\documentclass[prd,a4paper,superscriptaddress,nofootinbib,10pt]{revtex4} 
\usepackage{tikz}
\usetikzlibrary{decorations.markings}
\usetikzlibrary{decorations.pathmorphing}
  \usepackage[hidelinks]{hyperref}
\usepackage{graphicx}
\usepackage{amssymb}
\usepackage{amsmath}
\usepackage{lscape}
\usepackage{mathrsfs}
\usepackage{textcomp}
\usepackage{epsfig}
\usepackage{slashed}
\usepackage{xcolor}
\usepackage{subfigure}
\usepackage{bbold}
\usepackage{float}
\renewcommand{\d}{\mathrm{d}}

\begin{document}

\title{Fermion quasinormal modes on modified RN background}


\author{Nikola Herceg}
\email{Nikola.Herceg@irb.hr}
\affiliation{Rudjer Bo\v{s}kovi\'c Institute, Bijeni\v cka  c.54, HR-10002 Zagreb, Croatia}

\author{Nikola Konjik}
\email{konjik@ipb.ac.rs}
\affiliation{ Faculty of Physics, University of Belgrade,  Studentski trg 12, 11000 Beograd, Serbia}

\author{A. Naveena Kumara }
\email{nathith@irb.hr}
\affiliation{Rudjer Bo\v{s}kovi\'c Institute, Bijeni\v cka  c.54, HR-10002 Zagreb, Croatia}

\author{Andjelo Samsarov}
\email{asamsarov@irb.hr}
\affiliation{Rudjer Bo\v{s}kovi\'c Institute, Bijeni\v cka  c.54, HR-10002 Zagreb, Croatia}

\date{\today}

\begin{abstract}

Noncommutative (NC) geometry may open an alternative route  to quantum gravity. We study the influence of the spacetime noncommutativity  on the Dirac quasinormal modes in the modified Reissner-Nordstr\"om black hole spacetime.  The framework for the latter study is provided by a certain effective model of gravity coupled to fermions  which in itself encapsulates noncommutative deformation. This model describes a classical Dirac field coupled to a modified Reissner-Nordstr\"om geometry where the corresponding metric acquires an additional nonvanishing $r-\varphi $ component. As the earlier study shows,  this  model  appears to be equivalent to  a  model of semiclassical NC gauge theory  in which a NC gauge field is being coupled to a NC fermion field on the one side and the classical Reissner-Nordstr\"om background on the other. In comparison to the undeformed model where the Dirac field is coupled to the commutative Reissner-Nordstr\"om  black hole, the numerical results show that the oscillation frequencies and magnitude of damping of the Dirac quasinormal modes  change to an extent that cannot be neglected. In fact,   the influence of spacetime noncommutativity is shown to produce features reminiscent of a Zeeman-like splitting in the effective potential and quasinormal-mode spectrum.

\end{abstract}

\maketitle

\section{Introduction}

Black hole quasinormal modes (QNMs) are the characteristic frequencies of black holes in a ringdown phase calculated at linear order in perturbation theory. They appear as an outcome of a black hole’s  response  to an external perturbation and may be characterized by a superposition of exponentially damped oscillations with generally an infinite set of discrete frequencies and damping times. With the discovery of gravitational waves, the interest in their study increased considerably as they can be linked directly  and compared with  the experimentally observed  complex gravitational-wave frequencies of black hole merger remnants. The QNMs and their frequencies have a long history \cite{Regge:1957td,Zerilli:1970wzz,Zerilli:1970se,Vishveshwara:1970zz,Press:1971wr,Kokkotas:1999bd,Nollert:1999ji,Konoplya:2011qq}. They carry information about intrinsic properties of black holes.  Besides the characteristic parameters of black holes like mass, charge and angular momentum, they also give information about the stability of  black holes under perturbation by matter fields that evolve  in their  exterior region without backreacting on the metric. In general, the  frequencies themselves are complex, with the real part representing the oscillation frequency and the imaginary part describing the rate at which this oscillation is attenuated.  Accordingly, the stability of a black hole  is  guaranteed by the imaginary part  of QNMs being negative.

Rising excitement in QNMs, caused by experimental discovery of gravitational waves \cite{LIGOScientific:2016aoc}, also spurred the increased interest in the black hole spectroscopy which has been given a whole new set of possibilities. Indeed, quasinormal modes  are closely related to a  notion  of black hole spectroscopy which itself is based   on the correspondence between atomic and black hole spectra.  This correspondence has a root in  an analogy between the theory that describes scattering of gravitational waves on a black hole on the one side and the scattering theory in quantum mechanics on the other. It provides an appropriate frame   for  describing puzzling physical phenomena like Hawking radiation as a process of transitioning between different states in black hole QNM spectrum. This point of view  paves the way  to an intriguing connection between Hawking radiation and black hole quasinormal modes \cite{Corda:2012tz,Corda:2012dw,Corda:2013nza,Corda:2013paa,Konoplya:2004ik,Kiefer:2004ma}. Therefore, this correspondence by itself appears to be important in the route to quantizing gravity, because  one can naturally interpret black hole quasinormal modes in terms of quantum energy levels.

In this paper, we study the propagation and decay of  noncommutative massless fermionic fields in Reissner-Nordstr\"om (RN) black hole backgrounds  in order to document a possible appearance  of the fine structure in the spectrum, including some nonstandard features like the presence of anomalous decay rate behavior.  We carry out this study by using the method of continued fractions. It is worthy to mention that the study of Dirac QNMs has started with the paper by Cho, in which the massive and massless  Dirac quasinormal mode frequencies in the Schwarzschild black hole spacetime  have been calculated \cite{Cho:2003qe}. Refinement in the calculation of Dirac quasinormal modes in Schwarzschild background has been achieved by implementing the Leaver’s continued fraction method \cite{Leaver:1985ax} in the paper by Jing \cite{Jing:2005dt} and the high overtones of Dirac perturbations in the same black  hole setting  were studied in \cite{Castello-Branco:2004rzk}.  This analysis has been extended to the Reissner-Nordstr\"om background in \cite{Wu:2004vb}. Later on, QNMs of the Dirac field have been studied rather extensively for massless fermionic fields \cite{Zhidenko:2003wq,Jing:2003wq} and massive fermionic fields \cite{Chang:2006kf, Aragon:2020teq}  in Schwarzschild-de Sitter and Reissner-Nordstr\"om-de Sitter  backgrounds. Similarly, the Dirac QNMs of three-dimensional, four-dimensional and D-dimensional de Sitter spacetime have been respectively studied in \cite{Du:2004jt,Lopez-Ortega:2006tjo,Lopez-Ortega:2007vlo}. These results have shown  that fields with higher masses propagating in spacetimes with larger cosmological constants tend to decay more slowly in a Schwarzschild-de Sitter  black hole background \cite{Chang:2006kf}. Likewise, by using the convergent Frobenius  method  it has been shown that two chiralities of massive fermions give rise to an additional fine structure in the spectrum, for Schwarzschild and Kerr backgrounds \cite{Konoplya:2017tvu}.
Dirac QNMs have also been studied in other configurations of substantial interest \cite{Lopez-Ortega:2009jpx,Becar:2013qba,Lopez-Ortega:2010lld,Catalan:2013eza,Becar:2014jia,Stetsko:2016pau,Sakalli:2021dxd,Blazquez-Salcedo:2018bqy,Al-Badawi:2022aby}. 

The study of Dirac QNMs is essential for obtaining the  information about the stability of fermionic perturbations in the vicinity of a black hole and may help with providing insights into the behaviour of quantum fields in its background. In the  limit of large angular momentum, the QNMs start to nearly mirror/reflect the properties of photons being trapped into a closed circular orbits around black hole after they fall toward the horizon. These circular orbits are null  geodesics and they are usually referred to as light rings. For large angular momentum, i.e. in the eikonal limit, the orbital frequency of photons along light rings is closely given by the real part of the fundamental QNM frequency, while its imaginary part  closely corresponds to the Lyapunov exponent  \cite{Cardoso:2008bp,Konoplya:2017wot,Konoplya:2022gjp,Bolokhov:2023dxq}. The latter characterizes the stability of the  photon circular orbits along light rings. Quasinormal modes of fermion perturbations in the vicinity of black hole, as well as QNMs of other types of perturbations may be used to deduce these stability parameters related to null geodesics.
 
Moreover,  Dirac QNMs exhibit some unique features that distinguish them from scalar or gravitational perturbations, like for example  the spin-orbit coupling with the related effects as well as  a specific behaviour  with respect to the phenomenon of superradiance. Concerning the latter in particular, it is known that  the Dirac field behaves in a different way than the integer spin fields when they propagate in a curved background. For example, the fermion fields do not suffer from superradiant instabilities when they are scattered by rotating black holes \cite{Unruh:1974bw,Unruh:1973bda,Martellini:1977qf,Iyer:1978du,Ternov:1980st,Dolan:2015eua}. Indeed, while bosonic fields on  Kerr spacetime   become unstable in a certain regime of system parameters \cite{Lee:1977gk,Ternov:1978gq,Detweiler:1980uk,Rosa:2011my,Witek:2012tr,Pani:2012vp,Pani:2012bp}, the   fermionic  fields on  Kerr  spacetime  in that same regime remain stable  under condition of  extra  slow  rotation.

It is thus of interest to investigate QNM frequencies for the Dirac field in  spacetimes of various types. For the same reason, it is of interest to go one step further and consider Dirac field perturbations in spacetimes that incorporate essential/salient features of quantum gravity. This might help us resolve  some of the puzzles that surround the accurate theory of quantum gravity and bring us a few steps closer to its partial or even complete  understanding. Noncommutative spacetimes and the ensuing construction of gauge and gravity models   \cite{Seiberg:1999vs,Douglas:2001ba,Szabo:2001kg,Connes:1994yd,Ahluwalia:1993dd,Doplicher:1994zv,Doplicher:1994tu} on such manifolds are certainly able to incorporate some of the  salient features of quantum gravity and  meet the requirement of providing an effective model that is able to seize and describe its essence.

The premise of quantum nature of spacetime in the context of spacetime propagation of fermions and the associated equation of motion in the form of  Dirac equation    has been taken into account and considered  in a series of different studies \cite{Nowicki:1992if, Bertolami:2011rv, Kupriyanov:2013jka, Horvat:2011qn,Adorno:2009yu,Harikumar:2009wv,Andrade:2012pm,Horvat:2011iv,Buric:2010wd,Ching:2016nfv,Arzano:2021hpg,
Franchino-Vinas:2022fkh,Franchino-Vinas:2023rcc}. In particular,  quasinormal modes of the Dirac field have been studied within different approaches to effective quantum gravity, including models on noncommutative spacetime \cite{Gupta:2017lwk, Luo:2018tul}, as well as  certain models of  effective quantum gravity which arise when canonical quantum gravity leads to a semiclassical model described by an effective Hamiltonian constraint \cite{Malik:2024nhy, Zhang:2024khj}. Besides, it has been found that the issue of possible instability of the fermion field on the RN background due to  superradiant growth remains  unaffected by the  presence of spacetime noncommutativity. This appears to be a direct consequence of the weak energy condition not being violated by the NC deformation and the resulting perseverance  of the second law of black hole thermodynamics \cite{DimitrijevicCiric:2022ohs}.

In this work, we study  noncommutative massless Dirac field perturbations in a background of the RN black hole. They were  shown to be equivalent to considering the ordinary massless Dirac field perturbations in a background of  modified RN geometry, where the associated metric acquires an additional nonvanishing $r - \varphi$ component \cite{DimitrijevicCiric:2022ohs}. Furthermore, we  calculate the corresponding quasinormal mode frequencies using  the Leaver’s method of continued fractions followed by the accompanying  Gaussian  elimination procedure. The latter was devised with a purpose of bringing down the resultant $6$-term recurrence relations to the solvable $3$-term relations.  The quasinormal  modes of the Dirac field exhibit a discernable Zeeman-like splitting when the quantum parameter is introduced. As the  fundamental mode is the dominating one in the signal, only that mode will be the subject of our concern.

In Section \ref{preliminaries} we introduce the modified Reissner-Nordstr\"om metric with the $\star$-product of noncommutative spacetime and the associated tetrad field.
After selecting the appropriate ansatz for the Dirac equation, it reduces to two radial equations governing the two Weyl spinors. Using appropriate field redefinitions, the two equations can be condensed into a single second order ODE in the radial coordinate. In Section \ref{tortoise_sec} we recast the equation of motion into the Schr\"odinger form by introducing the tortoise coordinate and we identify the boundary conditions for QNMs.  We then set up the continued fraction method and solve the recurrence relations. Solutions of the recurrence relations are analyzed in Section \ref{results}, where we focus on the modifications of the QNM spectrum due to spacetime noncommutativity.
Closing remarks are given in Section \ref{conclusion}.


\section{Dirac equation in modified Reissner-Nordstr\"om spacetime} \label{preliminaries}

In references \cite{Ciric:2017rnf, DimitrijevicCiric:2019hqq}  a semiclassical model describing     charged NC scalar field $\hat{\Phi}$ and  NC $U(1)$ gauge field  $\hat{A}$ has been introduced. Both scalar and gauge NC fields interact with each other and they also interact with a classical  gravitational background of the RN type. The model is semiclassical in a sense that only gauge and scalar fields are considered to be affected  by the NC deformation, while on the other side, gravitational field is not. Instead, it is considered to be described by classical degrees of freedom, completely unaffected by the NC nature of spacetime. Model introduced in references \cite{Ciric:2017rnf, DimitrijevicCiric:2019hqq} therefore deals with a situation where the gauge and scalar fields are quantized and the gravitational field is not. It is however important to stress that the term "quantized", when applied to the gauge and scalar fields, refers here exclusively
to the situation when these fields are considered as noncommutative variables and does not refer to  the usual notion of second quantization in quantum field theory. 

The model was built using deformation quantization techniques based on Drinfeld twist operator and the explicit twist operator that was used in the construction was the so-called angular twist operator \cite{Ciric:2017rnf, DimitrijevicCiric:2019hqq}
\begin{eqnarray}  \label{AngTwist0Phi}
\mathcal{F} &=& e^{-\frac{i}{2}\theta ^{\alpha\beta}\partial_\alpha\otimes \partial_\beta} \nonumber\\
&=& e^{-\frac{ia}{2} (\partial_t\otimes\partial_\varphi - \partial_\varphi\otimes\partial_t)}, 
\end{eqnarray}
with $\alpha,\beta \in \{t,r, \theta, \varphi\}$ and $\theta^{t\varphi}= -\theta^{\varphi t}=a$  as the only non-zero components of the deformation tensor $\theta^{\alpha \beta}$.
Here, $a = 1 / \kappa$ is the deformation parameter that sets up the NC scale, commonly related to the Planck length.  This twist operator is constructed from Killing fields of the geometry that we consider -- it is a Killing twist. 

The $\star$-product, the wedge $\star$-product between forms, the coproduct and other structural maps of the related symmetry algebra can all be obtained from  the twist operator \eqref{AngTwist0Phi}. In particular, the $\star$-product between functions is given by
\begin{eqnarray}  \label{fStarg0Phi}
  f\star g &=&  \mu \big( e^{\frac{ia}{2} (\partial_t \otimes \partial_\varphi - \partial_\varphi \otimes  \partial_t)} f\otimes g \big) \nonumber\\
&=& fg + \frac{ia}{2}
(\partial_t f(\partial_\varphi g) - \partial_t g(\partial_\varphi f)) + 
\mathcal{O}(a^2) ,
\end{eqnarray}
where $\mu$ is the usual commutative pointwise multiplication of functions.
The remaining ingredients of the differential calculus are described in \cite{Ciric:2017rnf}.

It is noteworthy that some spacetime metrics may be deduced from  certain duality arguments, as demonstrated in \cite{DimitrijevicCiric:2022ohs}.
These arguments  are based on a recognition that in some cases twist deformed 
 $U(1)$ gauge theories on curved space
 behave in as much the same way as 
their commutative counterparts, albeit in a modified spacetime geometry.
In particular, in \cite{DimitrijevicCiric:2022ohs}  it has been shown that  the equation of motion  
 for a charged NC scalar field in a classical RN background coupled  to  NC $U(1)$ gauge field
may be rewritten in terms of the equation of motion governing the behaviour of a charged commutative scalar field, having the same charge $q$ as its NC counterpart,  and propagating in a modified RN geometry
\begin{equation} \label{NCdsRN}   
  {\rm d}s^2 = \Big(1-\frac{2MG}{r}+\frac{Q^2G}{r^2} \Big) {\rm d}t^2 - \frac{{\rm
d}r^2}{1-\frac{2MG}{r}+\frac{Q^2G}{r^2}} - aqQ \sin^2 \theta {\rm d} r {\rm d} \varphi - r^2({\rm d}\theta^2 + \sin^2\theta{\rm d}\varphi^2 ).
\end{equation}
It appears that  this novel,  first order effective  dual  metric (\ref{NCdsRN})  acquires an additional off-diagonal term which is induced purely by noncommutative nature of spacetime. This  feature  comes into play only in the presence of charged matter.

The geometry (\ref{NCdsRN})  can be considered as a noncommutative deformation of the   Reissner–Nordstr\"om (RN) metric
\begin{equation}
  {\rm d}s^2 = \Big(1-\frac{2MG}{r}+\frac{Q^2G}{r^2}\Big){\rm d}t^2 - \frac{{\rm
d}r^2}{1-\frac{2MG}{r}+\frac{Q^2G}{r^2}} - r^2({\rm d}\theta^2 + \sin^2\theta{\rm d}\varphi^2 )
. \label{dsRN}
\end{equation}
that represents a charged non-rotating black hole  with  mass $M$  and charge $Q$. For later purposes we shall introduce the usual abbreviation  $f= 1- \frac{2M}{r} + \frac{Q^2}{r^2}$.  Being static and spherically symmetric, the  spacetime of RN black hole
 has four Killing vectors, among which  $\partial_t$ and $\partial_\varphi $ are included,  and $t$ and $\varphi$ are the time and polar variables  
   of the spherical  coordinate system $\mu \in \{ t,r,\theta,\varphi\}$. 
It is now plainly seen that the twist (\ref{AngTwist0Phi})
 is a Killing twist,
as it is formed from the operators that are actually the Killing vectors for the metric (\ref{dsRN}).
Note that the implementation of the twist (\ref{AngTwist0Phi}) is compatible with the semiclassical nature of the model
that we consider bacause it ensures that  the geometry (\ref{dsRN})  remains unaffected by the deformation via this twist. This is because the twist (\ref{AngTwist0Phi}) does not act on the RN metric.


In this paper we consider a Dirac equation on the background geometry given by the modified  RN metric (\ref{NCdsRN}), where in addition we introduce a gauge potential $A_{\mu}$ minimally coupled to the Dirac operator $\gamma^a \nabla_a$,
\begin{equation}\label{diracnovogauge}
\big( i \gamma^a ( \nabla_a + A_a ) - m \big)\Psi = \big( i \gamma^a e_a^{~~\mu} ( \nabla_{\mu} + A_{\mu} ) - m \big)\Psi =0.
\end{equation}

Here the Latin index $a \in \{0,1,2,3 \}$ refers to the intrinsic coordinates and $\gamma^a$ are the standard flat space Dirac gamma matrices satisfying $\{ \gamma_a, \gamma_b \} = 2 \eta_{ab},$ where
\begin{equation} \label{metriceta}
  \eta_{ab} = \eta^{ab} =
\left( \begin{array}{ccccc}
  +1  & 0  & 0 & 0  \\
   0   & -1 & 0 & 0  \\ 
   0  & 0 & -1 &  0  \\
   0  & 0 & 0 & -1  \\
\end{array} \right).  
\end{equation}

The Dirac operator  $~  \gamma^a \nabla_a ~$ on a curved space
 is introduced in terms of tetrads (vierbeins) $~ e^a_{~\mu} ~$ and their inverse  $~ e_a^{~\mu}, ~$ satisfying
$~ e^a_{~\mu}  e_a^{~\nu} = \delta_{\mu}^{~\nu} ~$ and  $~ e^a_{~\mu}  e_b^{~\mu} = \delta^{a}_{~b}. ~$  Tetrads written in components are
$~  e^a_{~\mu} = ( e^a_{~t}, e^a_{~r}, e^a_{~\theta}, e^a_{~\varphi }) ~ $ and $ ~    e_a^{~\mu} = ( e_0^{~\mu}, e_1^{~\mu}, e_2^{~\mu}, e_3^{~\mu}). ~$ 
 They also satisfy 
$~g_{\mu \nu} =  e^a_{~\mu}  e^b_{~\nu}  \eta_{ab} ~$ and $~g^{\mu \nu} =  e_a^{~\mu}  e_b^{~\nu}  \eta^{ab}.$  
In what follows we use the setting defined in \cite{Dolan:2015eua}.
This setting consists of  the vierbein frame  chosen to be
\begin{equation} \label{metrictetrad}
  e^a_{~\mu} =
\left( \begin{array}{ccccc}
  \sqrt{f} & 0  & 0 & 0  \\
   0   & \frac{1}{\sqrt{f}} & 0 & 0  \\ 
   0  & 0 & r &  0 \\
 0  & \frac{aqQ }{2r} \sin \theta & 0 &  r \sin \theta \\
\end{array} \right)  \quad \quad \mbox{with the corresponding  inverse matrix}  \quad \quad
 e_a^{~\mu} =
\left( \begin{array}{ccccc}
  \frac{1}{\sqrt{f}} & 0  & 0 & 0 \\
   0   & \sqrt{f} & 0 &  -\frac{aqQ}{2 r^2} \sqrt{f}  \\ 
   0  & 0 &  \frac{1}{r} & 0 \\
  0  & 0 & 0 &  \frac{1}{r \sin \theta }  \\
\end{array} \right)
\end{equation}
and  the following representation of gamma matrices
\begin{equation} \label{gammarep}
  \gamma^0 = i  \tilde{\gamma}^0 =
 i \left( \begin{array}{ccccc}
  0  & I  \\
   I   & 0  \\ 
\end{array} \right),  \quad \quad 
 \gamma^1 =   i \tilde{\gamma}^3 =
i \left( \begin{array}{ccccc}
  0  & \sigma_3  \\
   -\sigma_3   & 0  \\  
 \end{array} \right), \quad \quad 
 \gamma^2 =   i \tilde{\gamma}^1 =
i\left( \begin{array}{ccccc}
  0  & \sigma_1  \\
   -\sigma_1   & 0  \\ 
\end{array} \right),  \quad \quad 
 \gamma^3 =   i \tilde{\gamma}^2 =
i\left( \begin{array}{ccccc}
  0  & \sigma_2  \\
   -\sigma_2   & 0  \\ 
\end{array} \right),  
\end{equation} 
where $\tilde{\gamma}^0$,  $\tilde{\gamma}^1$, $\tilde{\gamma}^2$ and $\tilde{\gamma}^3$ are gamma matrices in chiral/Weyl representation, while  $~ \sigma_i, ~ (i=1,2,3)~$ are the usual Pauli matrices,

\begin{equation} \label{Pauli}
   \sigma_1 =
\left( \begin{array}{ccccc}
  0  & 1  \\
   1   & 0  \\  
 \end{array} \right), \quad \quad 
 \sigma_2 =
\left( \begin{array}{ccccc}
  0  & -i  \\
   i   & 0  \\ 
\end{array} \right), \quad \quad 
  \sigma_3 =
\left( \begin{array}{ccccc}
  1  & 0  \\
   0   & -1  \\ 
\end{array} \right).  
\end{equation}

With the spinor field $\Psi$ written in terms of  two two-component spinors $\Psi_1$ and $\Psi_2$, namely $\Psi =
  (\Psi_1, \Psi_2)^T$ and the gauge potential $A_{\mu}   = (A_t, \vec{A}) = (-\frac{qQ}{r}, \vec{0}),$
the equation (\ref{diracnovogauge}) splits into two two-component equations
\begin{equation} \label{2compdirac}
\begin{split}
& \bigg[  - \frac{1}{\sqrt{f}} \mathbb{1} \partial_t   - \sqrt{f} \sigma_3 \partial_r - \frac{1}{2}  \frac{Mr - Q^2}{r^3} \frac{1}{\sqrt{f}} \sigma_3  - \frac{\sqrt{f}}{r} \sigma_3 -\frac{1}{r} \sigma_1 \partial_{\theta}   \\
& \quad + \frac{aqQ}{2r^2} \sqrt{f} \sigma_3 \partial_{\varphi}   - \frac{1}{r\sin \theta} \sigma_2 \partial_{\varphi}    
- \frac{1}{2r} \cot \theta \sigma_1   - \frac{iqQ}{r \sqrt{f}} \mathbb{1}      \bigg] \Psi_2 -  m \mathbb{1}  \Psi_1 =0, \\ \\
& \bigg[- \frac{1}{\sqrt{f}} \mathbb{1} \partial_t +   \frac{1}{2}  \frac{Mr - Q^2}{r^3} \frac{1}{\sqrt{f}} \sigma_3 +
     \sqrt{f} \sigma_3 \partial_r  + \frac{\sqrt{f}}{r} \sigma_3 + \frac{1}{r} \sigma_1 \partial_{\theta}   \\
&\quad - \frac{aqQ}{2r^2} \sqrt{f} \sigma_3 \partial_{\varphi}
+ \frac{1}{r\sin \theta} \sigma_2 \partial_{\varphi}  
   + \frac{1}{2r} \cot \theta \sigma_1  - \frac{iqQ}{r \sqrt{f}} \mathbb{1}   \bigg] \Psi_1 - m \mathbb{1}  \Psi_2 =0.
\end{split}
\end{equation}
For details, we refer the reader to the reference  \cite{DimitrijevicCiric:2022ohs}. In order to separate this system of equations, we  impose the following ansatz  
\begin{equation} \label{anz1}
\Psi =
  e^{i(\nu \varphi - \omega t)}   \left(\begin{matrix} \psi_1 (r, \theta)  \\ \psi_2 (r, \theta )  \\ \end{matrix}\right)  =
   e^{i(\nu \varphi - \omega t)}  \left(\begin{matrix}     p R_2(r) S_1(\theta)  \\ - \frac{1}{r} R_1(r) S_2(\theta)  \\   \frac{1}{r} R_1(r) S_1(\theta)  \\ R_2(r) S_2(\theta)  \\ \end{matrix}\right).
\end{equation}
In the above ansatz $p$ is a free parameter which can acquire two possible values, $p\in\{-1, +1 \}$. We keep it   in order to be able to cover and analyse
a somewhat more general scope of possibilities.

After inserting the ansatz  (\ref{anz1}) into (\ref{2compdirac}) and dividing the resulting equations  respectively with $ R_2 S_1 $, $R_1 S_2$, $R_1 S_1$, $R_2 S_2$, one finds that this system of equations  is completely separable,
\begin{equation} \label{anz2}
\begin{split}
&\frac{i \omega }{\sqrt{f}}   \frac{R_1}{R_2} - r\sqrt{f}   \frac{1}{R_2} \partial_r \Big( \frac{R_1}{r} \Big) - \frac{1}{2} \frac{Mr - Q^2}{r^3} \frac{1}{\sqrt{f}} \frac{R_1}{R_2}
   -  \frac{\sqrt{f}}{r} \frac{ R_1 }{R_2} +  i \nu \frac{aqQ}{2r^2} \sqrt{f}  \frac{R_1}{R_2}  
- \frac{iqQ}{r \sqrt{f}} \frac{R_1}{R_2} - p  mr    \\ 
&  \quad \quad \quad \quad = \frac{\partial_{\theta} S_2}{S_1}  + \frac{ \nu }{\sin \theta} \frac{ S_2}{S_1}   + \frac{1}{2} \cot \theta \frac{ S_2}{S_1} \equiv \lambda,\\ \\
&\frac{i \omega r^2}{\sqrt{f}} \frac{R_2}{R_1}   +  r^2 \sqrt{f} \frac{ 1 }{R_1} \partial_r R_2 
+\frac{1}{2} \frac{Mr - Q^2}{r} \frac{1}{\sqrt{f}} \frac{R_2}{R_1} + r\sqrt{f} \frac{R_2}{R_1}
-  i \nu \frac{aqQ}{2} \sqrt{f}  \frac{R_2}{R_1}  
- \frac{iqQ r}{ \sqrt{f}} \frac{R_2}{R_1}  +  mr \\
&  \quad \quad \quad \quad = \frac{\partial_{\theta} S_1}{S_2}  - \frac{ \nu }{\sin \theta} \frac{ S_1}{S_2}   + \frac{1}{2} \cot \theta \frac{ S_1}{S_2} \equiv \lambda_1, \\ \\
& p\frac{i \omega r^2}{\sqrt{f}} \frac{R_2}{R_1}   +  r^2 \sqrt{f} \frac{ p }{R_1} \partial_r R_2 
+\frac{1}{2} \frac{Mr - Q^2}{r} \frac{p}{\sqrt{f}} \frac{R_2}{R_1} + p r\sqrt{f} \frac{R_2}{R_1}
- p i \nu \frac{aqQ}{2} \sqrt{f}  \frac{R_2}{R_1}  
- p\frac{iqQ r}{ \sqrt{f}} \frac{R_2}{R_1}  -  mr \\
&  \quad \quad \quad \quad = \frac{\partial_{\theta} S_2}{S_1}  + \frac{ \nu }{\sin \theta} \frac{ S_2}{S_1}   + \frac{1}{2} \cot \theta \frac{ S_2}{S_1} = \lambda, \\ \\
& -\frac{i \omega }{\sqrt{f}}   \frac{R_1}{R_2} + r\sqrt{f}   \frac{1}{R_2} \partial_r \Big( \frac{R_1}{r} \Big) + \frac{1}{2} \frac{Mr - Q^2}{r^3} \frac{1}{\sqrt{f}} \frac{R_1}{R_2}
   +  \frac{\sqrt{f}}{r} \frac{ R_1 }{R_2} -  i \nu \frac{aqQ}{2r^2} \sqrt{f}  \frac{R_1}{R_2}  
+ \frac{iqQ}{r \sqrt{f}} \frac{R_1}{R_2} -   mr    \\  
&  \quad \quad \quad \quad = -p\frac{\partial_{\theta} S_1}{S_2}  + p\frac{ \nu }{\sin \theta} \frac{ S_1}{S_2}   - p \frac{1}{2} \cot \theta \frac{ S_1}{S_2} = -p \Big(   \frac{\partial_{\theta} S_1}{S_2}  -\frac{ \nu }{\sin \theta} \frac{ S_1}{S_2}   + \frac{1}{2} \cot \theta \frac{ S_1}{S_2}    \Big)  = - p \lambda_1.
\end{split}
\end{equation}
 It is easily seen that the two separation constants $\lambda$ and $\lambda_1$, which have appeared in the process of separation, are not mutually  independent, but subject to a certain constraint, as we shall see shortly.  
Moreover, it appears that a separation of the system (\ref{2compdirac}) can be  achieved for both values of $p$. Let us now analyse different possibilities as related to the choice of $p$ and the  implications this choice has for the relation between the separation constants $\lambda$ and $\lambda_1$.

For that purpose, we note that while the match between the first and the fourth equation in (\ref{anz2}) leads to the condition  $\lambda + pmr = p\lambda_1 - mr,$ the match between the second and the third yields
$\lambda_1 - mr = p\lambda + p mr.$ These two conditions are in fact  equivalent to each other, as  can readily be seen by multiplying any of them by $p$.
For $p=-1,$ two separation constants $\lambda$ and $\lambda_1$ can be related and the relation among them is given by $\lambda_1 = -\lambda$. On the opposite, for $p=+1,$
the only viable possibility is attained when $m=0$\footnote{For $p=+1$ the above obtained condition leads to $\lambda_1 = \lambda + 2mr, $  which is inconsistent with $\lambda$ and $\lambda_1$ both being the separation constants (independent of $r$ and $\theta$).}. Although 
 the system (\ref{2compdirac}) is still separable for $p=+1$,  the separation constants $\lambda$ and $\lambda_1$ can be related only if the spinor field is massless,
in which case the corresponding relation is $\lambda_1 = \lambda$.

We can conclude that for the spinor field with  nonzero mass, the parameter $p$  must necessarily be equal to $p=-1$. On the contrary, for massless field, the parameter $p$
may take any of the two possible values, $p=+1$ or $p=-1.$ Generally, we can write $\lambda_1 = p\lambda$ with a remark that for $p=+1$, the mass of the field must be zero.
For $p=-1$ the spinor field may be massive, as well as massless.

 In effect, the system of equations (\ref{2compdirac}) gives rise to two independent  angular equations
\begin{equation} \label{anz3}
 \begin{split}  
&\partial_{\theta} S_2 + \frac{\nu}{\sin \theta} S_2 + \frac{1}{2} \cot \theta S_2 = \lambda S_1,  \\
& \partial_{\theta} S_1 - \frac{\nu}{\sin \theta} S_1 + \frac{1}{2} \cot \theta S_1 = \lambda_1 S_2,
\end{split}
\end{equation}
 and two independent radial equations
\begin{equation} \label{anz4}
\begin{split}
\frac{i \omega }{\sqrt{f}}   R_1 - \sqrt{f}   \partial_r R_1  - \frac{1}{2} \frac{Mr - Q^2}{r^3} \frac{1}{\sqrt{f}} R_1
    +  i \nu \frac{aqQ}{2r^2} \sqrt{f}  R_1 
- \frac{iqQ}{r \sqrt{f}} R_1  
 = (\lambda + pmr) R_2,     \\
\frac{i \omega r^2 }{\sqrt{f}}   R_2 +  r^2 \sqrt{f}   \partial_r R_2  + \frac{1}{2} \frac{Mr - Q^2}{r} \frac{1}{\sqrt{f}} R_2 + r \sqrt{f} R_2
    -  i \nu \frac{aqQ}{2} \sqrt{f}  R_2 
- \frac{iqQr}{ \sqrt{f}} R_2  
 = (p\lambda - mr) R_1
\end{split}
\end{equation}
This system of radial equations can be used to study the behaviour of fermion quasinormal modes in the modified  RN background (\ref{NCdsRN}).

Upon decoupling, these two equations give rise to the following two 2nd order differential equations,
\begin{equation} \label{anz5}
\begin{split}
-r^2 f \partial_r^2 R_1 + \Bigg( M-r - \frac{Mr - Q^2}{r}  + i \nu a qQ f + \frac{pmr^2}{\lambda + pmr} f \Bigg) \partial_r R_1   
+ \Bigg[  \frac{Mr - Q^2}{4r} \frac{\partial_r f}{f}  - \frac{1}{2} i\omega r^2   \frac{\partial_r f}{f}  
 -   \frac{i \omega pmr^2}{\lambda + pmr} \\
  -\frac{1}{2} \frac{-2 \lambda M r^3 - 3pm M r^4 + 3\lambda Q^2 r^2  + 4pm Q^2 r^3}{r^4 (\lambda + pmr)} +   \frac{i \nu aqQ}{4}  \partial_r f 
-  \frac{i \nu aqQ}{2}   f   \frac{2 \lambda r + 3pm r^2}{r^2 (\lambda + pmr)} + \frac{1}{2} iqQ r  \frac{\partial_r f}{f} \\
   + iqQ   \frac{ \lambda  + 2pm r}{\lambda + pmr}  
+ \Big(i \omega r^2 + \frac{1}{2} \frac{Mr - Q^2}{r} + rf -   \frac{i \nu aqQ}{2}  f - iqQr  \Big) \Big(  \frac{i\omega}{f} - \frac{1}{2f} \frac{Mr - Q^2}{r^3}
    +   \frac{i \nu aqQ}{2 r^2}  - \frac{iqQ}{r} \frac{1}{f}    \Big) \Bigg] R_1   \\
 = (\lambda + pmr)(p\lambda - mr) R_1,     ~~~~~~~~~~~~~~~~~~~~~~~~~~~~~~~~~~~~~~~~~~~~~~~~~~~~~~~~~~~~~~~~~~~~~~~~~~~~~~~~
\end{split}
\end{equation}
for the first component  $R_1 \equiv R_{s=-\frac{1}{2}}$ and
\begin{equation} \label{anz6}
\begin{split}
-r^2 f \partial_r^2 R_2 + \Bigg( 5M- 3r - \frac{2Q^2}{r}- \frac{Mr - Q^2}{r}  + i \nu a qQ f - \frac{mr^2}{p\lambda - mr} f \Bigg) \partial_r R_2  
\qquad \qquad \qquad \qquad \qquad   \qquad \qquad \qquad \qquad \qquad  \qquad \qquad \qquad \qquad \qquad \\
+ \Bigg[   \Big(i \omega r^2 + \frac{1}{2} \frac{Mr - Q^2}{r} + rf -   \frac{i \nu aqQ}{2}  f - iqQr  \Big) \Big(  \frac{i\omega}{f} - \frac{1}{2f} \frac{Mr - Q^2}{r^3}
    +   \frac{i \nu aqQ}{2 r^2}  - \frac{iqQ}{r} \frac{1}{f}  + \frac{1}{2} \frac{\partial_r f}{f} - \frac{m}{p\lambda - mr}  \Big)   \qquad \qquad \qquad  \qquad \qquad \qquad \qquad \\
 -  2i \omega r - \frac{1}{2} \frac{Q^2}{r^2} - f - r\partial_r f + \frac{i \nu aqQ}{2} \partial_r f  + iqQ  \Bigg] R_2   
 = (p\lambda - mr)(\lambda + p mr) R_2,        ~~~~~~~~~~~~~~~~~~~~~~~~~~~~~~~~~~~~~~~~~~~~~~~~~~~~~~~~~~~~~~~~~~~~~~~~~~~~~~~~
\end{split}
\end{equation}
for the second component  $R_2 \equiv R_{s=+\frac{1}{2}}$, where $s$ refers to chirality\footnote{For massless particles chirality turns out to be the same as helicity. However, for massive particles these two notions  need to be distinguished.}. Separating the right handed component of the Dirac spinor from the left handed one implies that $\lambda$ satisfies $\lambda^2 = (j-s)(j+s+1)$ \cite{Dolan:2015eua}. Interestingly, if we take the notation $\lambda \equiv \lambda_s,$  where $s \in \{ -1/2, +1/2 \},$ then it is straightforward to see that  $\lambda^2_{-\frac{1}{2}} = \lambda^2_{+\frac{1}{2}} +1.$  We shall make use of this  relation in the subsequent analysis
 when trying to relate  the separation constants appearing in the two component equations (\ref{anz7}) and (\ref{anz8}) down below.   At this point, it is also convenient to introduce the label $\Delta = r^2 f.$

As a next step, note that if the transformation $R_1 \equiv R_{s=-\frac{1}{2}} = r^{\beta} \Delta^{\alpha/2} \xi_{s=-\frac{1}{2}}$ is carried out on the equation (\ref{anz5}), then for $\beta = \frac{1}{2}$ and $\alpha \equiv s = -\frac{1}{2},$ this equation transforms into
\begin{eqnarray}   \label{anz7}
 \Delta \partial_r^2 \xi_{-\frac{1}{2}} &+& \Bigg(  2(\alpha + 1) (r - M)  - i \nu a qQ f - \frac{pmr^2}{\lambda + pmr} f \Bigg) \partial_r \xi_{-\frac{1}{2}}  \nonumber \\
&+&\Bigg[ \frac{{( \omega r^2 - qQr )}^2  + i(r - M)( \omega r^2 - qQr ) }{\Delta} +iqQ - 2i \omega r + p \lambda^2  \Bigg] \xi_{-\frac{1}{2}} \\
&-&\Bigg[ \frac{i\nu a qQ }{r^3}  \Big(  \alpha r^2  +  (1- \alpha) Mr  - Q^2 \Big) + \frac{pm}{\lambda + pmr} \Bigg( (\alpha + \frac{1}{2}) (r - M) - i \omega r^2 + iqQr - \frac{i\nu aqQ}{2} f   \Bigg)  + pm^2 r^2  \Bigg] \xi_{-\frac{1}{2}}  =0.    \nonumber
\end{eqnarray}
Likewise, if the same transformation $R_2 \equiv R_{s=\frac{1}{2}} = r^{\beta} \Delta^{\alpha/2} \xi_{s=\frac{1}{2}}$ is applied to the equation (\ref{anz6}) for the  component $R_2$, then the choice $\beta = -\frac{1}{2}$ and $\alpha \equiv s = \frac{1}{2}$ yields\footnote{At this point, $\alpha \equiv s$ is a generic parameter which may take one of two values, $\alpha \equiv s = -\frac{1}{2}$  or $\alpha \equiv s = +\frac{1}{2}.$ Which one of these two values  will it take depends on the component it refers to, as well as the equation in which it appears. Therefore,
if the parameter $s$ appears in the  equation (\ref{anz7}) and refers to the first component $R_1 \equiv R_s,$ it will acquire the value $-\frac{1}{2}$, and when it appears
 in the  equation (\ref{anz8}) that  governs behaviour of the second component $R_2 \equiv R_s,$ it will take the value $+\frac{1}{2}$. Note that in all terms except the middle one, we have retained a generic label $\alpha,$ avoiding to fix it to a particular value, so that later on we can more easily deduce a generic form of the 2nd order differential equation that would encompass both components within a single equation.}
 
\begin{eqnarray}   \label{anz8}
 \Delta \partial_r^2 \xi_{+\frac{1}{2}} &+& \Bigg(  2(\alpha + 1) (r - M)  - i \nu a qQ f + \frac{mr^2}{p\lambda - mr} f \Bigg) \partial_r \xi_{+\frac{1}{2}}  \nonumber \\
&+&\Bigg[ \frac{{( \omega r^2 - qQr )}^2  - i(r - M)( \omega r^2 - qQr ) }{\Delta} - iqQ + 2i \omega r + p \lambda^2 + 1 \Bigg] \xi_{+\frac{1}{2}} \\
&-&\Bigg[ \frac{i\nu a qQ }{r^3}  \Big(  \alpha r^2  +  (1- \alpha) Mr  - Q^2 \Big) - \frac{m}{p\lambda - mr} \Bigg( (\alpha + \frac{1}{2}) (r - M) + i \omega r^2 - iqQr - \frac{i\nu aqQ}{2} f   \Bigg)  + pm^2 r^2  \Bigg] \xi_{+\frac{1}{2}}  =0.    \nonumber
\end{eqnarray}
We point out that  in the limit $m, a \to 0$ the above equations  both  reduce to the equations studied in \cite{Richartz:2014jla}.

From now on, we  put the parameter $p=-1$ in the above 2nd order equations. This is because   it is the only viable choice in the  case of the  spinor field with mass and it is also valid
in the case of massless spinor field.  Setting $p=-1$  leads to
\begin{eqnarray}   \label{anz9}
 \Delta \partial_r^2 \xi_{-\frac{1}{2}} &+& \Bigg(  2(\alpha + 1) (r - M)  - i \nu a qQ f + \frac{m\Delta}{\lambda - mr}  \Bigg) \partial_r \xi_{-\frac{1}{2}}  \nonumber \\
&+&\Bigg[ \frac{{( \omega r^2 - qQr )}^2  + i(r - M)( \omega r^2 - qQr ) }{\Delta} +iqQ - 2i \omega r  - \lambda^2  \Bigg] \xi_{-\frac{1}{2}} \\
&-&\Bigg[ \frac{i\nu a qQ }{r^3}  \Big(  \alpha r^2  +  (1- \alpha) Mr  - Q^2 \Big) - \frac{m}{\lambda - mr} \Bigg( (\alpha + \frac{1}{2}) (r - M) - i \omega r^2 + iqQr - \frac{i\nu aqQ}{2} f   \Bigg)  - m^2 r^2  \Bigg] \xi_{-\frac{1}{2}}  =0    \nonumber
\end{eqnarray}

\begin{eqnarray}   \label{anz10}
 \Delta \partial_r^2 \xi_{+\frac{1}{2}} &+& \Bigg(  2(\alpha + 1) (r - M)  - i \nu a qQ f - \frac{m\Delta}{\lambda + mr}  \Bigg) \partial_r \xi_{+\frac{1}{2}}  \nonumber \\
&+&\Bigg[ \frac{{( \omega r^2 - qQr )}^2  - i(r - M)( \omega r^2 - qQr ) }{\Delta} - iqQ + 2i \omega r - \lambda^2 + 1 \Bigg] \xi_{+\frac{1}{2}} \\
&-&\Bigg[ \frac{i\nu a qQ }{r^3}  \Big(  \alpha r^2  +  (1- \alpha) Mr  - Q^2 \Big) + \frac{m}{\lambda + mr} \Bigg( \big(\alpha + \frac{1}{2}\big) (r - M) + i \omega r^2 - iqQr - \frac{i\nu aqQ}{2} f   \Bigg)  - m^2 r^2  \Bigg] \xi_{+\frac{1}{2}}  =0    \nonumber
\end{eqnarray}

The transformation $R_s = r^{-s} \Delta^{s/2} \xi_s$ has a generic form that enables  both components $R_1$ and $R_2$ to be factorised at once. We want to achieve a similar thing
with the system of two 2nd order differential equations  (\ref{anz9}) and (\ref{anz10}).  More precisely, we want to write a generic form of the equation that would unify both equations 
(\ref{anz9}) and (\ref{anz10})  into a single equation valid for both components $\xi_{-\frac{1}{2}}$
and $\xi_{+\frac{1}{2}}$. This generalization is  straightforward  and  is  achieved  by unification in the form
\begin{eqnarray}   \label{anz11}
 \Delta \partial_r^2 \xi_s &+& \Bigg(  2(s + 1) (r - M)  - i \nu a qQ f -  2s \frac{m\Delta}{\lambda_s + 2s mr}  \Bigg) \partial_r \xi_s  \nonumber \\
&+&\Bigg[ \frac{{( \omega r^2 - qQr )}^2  - 2 i s(r - M)( \omega r^2 - qQr ) }{\Delta} + 4i s \omega r - 2 i s qQ  - \lambda_s^2  \Bigg] \xi_s \\
&-&\Bigg[ \frac{i\nu a qQ }{r^3}  \Big(  s r^2  +  (1- s) Mr  - Q^2 \Big) + \frac{m}{\lambda_s + 2s mr} \Bigg(2s (s + \frac{1}{2}) (r - M) + i \omega r^2 - iqQr - 2s\frac{i\nu aqQ}{2} f   \Bigg)  - m^2 r^2  \Bigg] \xi_s  =0.    \nonumber
\end{eqnarray}
Here we took  the notation $\lambda \equiv \lambda_s,$  where $s \in \{ -1/2, +1/2 \},$ as explained earlier,
and used the identity $\lambda^2_{-\frac{1}{2}} = \lambda^2_{+\frac{1}{2}} +1$.  

This is the required equation that has  the right commutative  limit coinciding with the known results in the literature \cite{Richartz:2014jla}. Moreover, from  this equation the noncommutativity contribution, as well as the field mass contribution can be precisely isolated.
In order to proceed further in complete generality, one needs to find the appropriate tortoise coordinate, which again in the limit $a, m \rightarrow0$  reduces to the 
known tortoise coordinate for RN. However, we shall not pursue the analysis further in its whole generality and in the rest of the paper will  rather focus
on massless, but noncommutative fermion perturbations.
We will address the massive case in a separate study.

For the end of this section, note that we could arrive at the equation  (\ref{anz11})  directly from (\ref{2compdirac}) by assuming the following ansatz from the very beginning,
\begin{equation} \label{anz12}
\Psi =
  e^{i(\nu \varphi - \omega t)}   \left(\begin{matrix} \psi_1 (r, \theta)  \\ \psi_2 (r, \theta )  \\ \end{matrix}\right)  =
   e^{i(\nu \varphi - \omega t)}  \left(\begin{matrix}     p r^{-1/2} \Delta^{1/4} \xi_{+\frac{1}{2}} (r) S_1 (\theta) \\- r^{-1/2} \Delta^{-1/4} \xi_{-\frac{1}{2}} (r) S_2 (\theta)  \\  
     r^{-1/2} \Delta^{-1/4} \xi_{-\frac{1}{2}} (r) S_1 (\theta)  \\ r^{-1/2} \Delta^{1/4} \xi_{+\frac{1}{2}} (r) S_2 (\theta)  \\ \end{matrix}\right).
\end{equation}

\section{Tortoise coordinate and the boundary conditions} \label{tortoise_sec}

\subsection{The tortoise coordinate and the effective potential}

Let us focus on the situation with vanishing mass, $m=0.$
Then  the tortoise coordinate $y$ can be introduced in which the radial equation (\ref{anz11})
takes the Schr\"odinger form
\begin{equation} \label{schrod2}
\frac{\d^2 \chi}{\d {y}^2}  + V \chi =0,
\end{equation}
and  which is defined by the following change of coordinates
\begin{equation} \label{modtortoise}
\frac{\d y}{ \d r} = \frac{1}{f \bigg( 1+ ia\nu \frac{qQ}{r} \bigg)}.
\end{equation}
Integrating (\ref{modtortoise}) we find
\begin{eqnarray} \label{modtortoise1}
y &=& y^{(0)} -ia \nu qQ  ~ \Bigg \{  \frac{r_+}{r_+ - r_-} \ln (r- r_+) - \frac{r_-}{r_+ - r_-} \ln (r- r_-) \Bigg \}  \\
&=& r + \frac{r_+}{r_+ - r_-} \Big(r_+ - ia\nu qQ \Big) \ln (r- r_+) - \frac{r_-}{r_+ - r_-} \Big(r_- - ia\nu qQ \Big) \ln (r- r_-)  ,\nonumber
\end{eqnarray}
where $y^{(0)}$ is the standard tortoise coordinate for the Reissner–Nordstr\" om metric given by
\begin{equation}  
y^{(0)} \equiv r_*^{RN} = r + \frac{r^2_+}{r_+ - r_-} \ln (r- r_+) - \frac{r^2_-}{r_+ - r_-} \ln (r- r_-). \label{StandardTortCoord}
\end{equation}

More precisely, for $m=0$ the transformation  $\chi_s (r) = \Delta^{s/2} r \xi_s (r) $ followed by  the change of coordinate  (\ref{modtortoise}),
brings the equation (\ref{anz11}) into the form
\begin{eqnarray}   \label{modtort2}
  \frac{\partial^2 \chi_s }{\partial y^2 } &+& \frac{\Delta}{r^4} \Bigg[ \frac{2Q^2}{r^2} - \frac{2M}{r} - j(j+1) + s^2 
    +\frac{{  \big( \omega r^2 - qQr  - i s(r - M) \big) }^2   }{\Delta} + 4i s \omega r - 2 i s qQ
 + \frac{ i a \nu  qQ \Delta}{r^3}      \nonumber \\
&+&    is a \nu qQ \frac{r - M}{r^2}  - \frac{ i a \nu  qQ }{r^3}  \bigg( s r^2 + (1-s) Mr - Q^2 \bigg)  + 2i a \nu \frac{qQ}{r} \Big(  \frac{2Q^2}{r^2} - \frac{2M}{r} - j(j+1) + s^2   \Big)  
      \\
&+& 2i a \nu \frac{qQ}{r} \frac{{ \big( \omega r^2 - qQr  - i s(r - M) \big) }^2   }{\Delta}   - 8s a \nu  \omega qQ + 4 s a \nu  \frac{q^2 Q^2}{r} \Bigg] \chi_s  =0.    \nonumber
\end{eqnarray}
It has  the form of the equation (\ref{schrod2}), with the potential being given by
\begin{eqnarray} \label{veffektive}
V &=& \frac{\Delta}{r^4} \Bigg[ \frac{2Q^2}{r^2} - \frac{2M}{r} - j(j+1) + s^2 
    +\frac{{  \big( \omega r^2 - qQr  - i s(r - M) \big) }^2   }{\Delta} + 4i s \omega r - 2 i s qQ
 + \frac{ i a \nu  qQ \Delta}{r^3}      \nonumber \\
&+&    is a \nu qQ \frac{r - M}{r^2}  - \frac{ i a \nu  qQ }{r^3}  \bigg( s r^2 + (1-s) Mr - Q^2 \bigg)  + 2i a \nu \frac{qQ}{r} \Big(  \frac{2Q^2}{r^2} - \frac{2M}{r} - j(j+1) + s^2   \Big)  
      \\
&+& 2i a \nu \frac{qQ}{r} \frac{{ \big( \omega r^2 - qQr  - i s(r - M) \big) }^2   }{\Delta}   - 8s a \nu  \omega qQ + 4 s a \nu  \frac{q^2 Q^2}{r} \Bigg].    \nonumber
\end{eqnarray}
We point out that this equation is valid up to first order in the deformation parameter $a$. Note also that  (\ref{modtortoise}) is the same change of coordinate
that was used in  \cite{DimitrijevicCiric:2019hqq} in order to accomplish the same task of retrieving the effective potential in the context of NC scalar perturbations.

\subsection{Boundary conditions}

In general, the boundary conditions impose the constraints on the shape of the solutions 
in the asymptotic regions of space. In order to deduce the appropriate boundary conditions for QNM spectrum of fermion perturbations, we need to consider the limiting forms of the
equation  (\ref{modtort2})  in two separate asymptotic regions of space, the spatial infinity $r\to \infty$ and  the region near the event horizon $r\to r_+$. 
The required limiting forms of the solution
are obtained by solving equation  (\ref{modtort2})  analytically in two stated regions of space and then
 imposing the QNM boundary condition of purely incoming waves on the horizon and purely outgoing waves at the infinity. 

We first consider  the $r \rightarrow \infty$  limit. The equation  (\ref{modtort2}) in this limit reduces to
\begin{equation} \label{modtort3}
\frac{ \d^2 \chi_s}{ \d y^2} + \Big[  \omega^2  - 2 \frac{ \omega qQ }{y} + 2is \frac{\omega }{y} + 2ia\nu  \frac{qQ}{y}        
   \omega^2   \Big] \chi_s =0. 
\end{equation}
Note that in this limit the tortoise coordinate does not actually differ from the radial coordinate.
The solution to the equation (\ref{modtort3}) is given by
\begin{equation} \label{modtort4}
\xi_s = \frac{\chi_s}{r \Delta^{s/2}}  \sim e^{\pm i \omega y} y^{-1-i  qQ - 2s - a\nu qQ \omega}. 
\end{equation}
 Likewise, in the near horizon limit $r \rightarrow r_+,$ the
 equation (\ref{modtort2}) boils down to
\begin{equation} \label{modtort5}
\frac{ \d^2 \chi_s}{ \d y^2} +  \frac{1}{r_+^4}
\Big(  \omega r_+^2 -  qQ  r_+   - is(r_+ - M)\Big)^2  \Big( 1+ 2ia\nu \frac{qQ}{r_+}  \Big) \chi_s =0, 
\end{equation}
with the solution
\begin{equation} \label{modtort6}
\xi_s = \frac{\chi_s}{r \Delta^{s/2}}   \sim  \frac{1}{{(r - r_+)}^{s/2}} e^{\pm i \big( \omega  - \frac{ qQ}{r_+}  - is \frac{r_+ - r_-}{2r_+^2} \big)  \big( 1 + ia\nu  \frac{ qQ}{r_+}  \big)y }. 
\end{equation}
It should be stressed that the solutions (\ref{modtort4}) and (\ref{modtort6}) are perturbative in the NC parameter $a$ and are valid only up to first order in $a$. The QNM boundary conditions, purely outgoing in the infinity and purely incoming at the horizon, select signs in (\ref{modtort4}) and (\ref{modtort6}). Consequently, the asymptotic form of the quasinormal mode
solutions  can be summarized as
\begin{gather}    
  \xi_s (r) \rightarrow 
\begin{cases}         
  Z_{\text{out}} e^{i \omega y} y^{-1-i  qQ  - 2s -  a\nu qQ \omega}, & \text{for } r \rightarrow \infty,\>\> (y\rightarrow \infty) \\  
  \\
  Z_{\text{in}} \frac{1}{{(r - r_+)}^{s/2}}  e^{-i \Big( \omega  - \frac{ qQ}{r_+} - is \frac{r_+ - r_-}{2r_+^2} \Big)  \Big( 1 + ia \nu  \frac{ qQ}{r_+}  \Big)y },   & \text{for }r \rightarrow  r_+, \>\> (y\rightarrow -\infty)                
\end{cases} . \label{ncboundaryconditions}   
\end{gather}
Here $Z_{\text{out}}$ and  $Z_{\text{in}}$ are the amplitudes of the outgoing and ingoing waves, respectively, and they do not depend on $r$ (or $y$). In the special case of   vanishing spacetime deformation $a=0$, these asymptotic solutions
reduce to the asymptotic solutions of \cite{Richartz:2014jla}. 

Equation (\ref{anz11}) has an irregular singularity at $r=+\infty$ and three regular singularities at
$r=0$, $r=r_-$ and $r=r_+$. In order to apply  the Leaver's method of continued fractions, we expand the general solution of (\ref{anz11}) in terms of powers series
around  $r = r_+$. Then the radial part of the spin $1/2$  field takes the form 
\begin{equation}  \label{generalpowersolution}
 \xi_s (r) = e^{i \omega r}  {(r-r_-)}^{\epsilon} \sum_{n=0}^{\infty} a_n {\Big( \frac{r-r_+}{r-r_-} \Big)}^{n + \delta}.
\end{equation}

At this stage, the parameters $\delta$ and $\epsilon$ are still unknown. However, they can be fixed by demanding that the solution (\ref{generalpowersolution}) satisfies the  boundary conditions (\ref{ncboundaryconditions}) at the horizon and at the infinity.
From  the general   solution (\ref{generalpowersolution}) to the equation (\ref{anz11}), it is possible to  infer  its most prevailing behaviour in both critical regimes, that close to the event horizon, as well as that at far infinity. 
 While the dominant behaviour   of   (\ref{generalpowersolution})   in the regime $r \rightarrow \infty$  is governed by the term
$ r^{\epsilon} e^{i \omega r}, $  its dominant behaviour  in the regime   $r \rightarrow r_+$   is determined by the term $ {( r - r_+ )}^{\delta} $. Moreover, an identification of the leading contributions to the general solution (\ref{generalpowersolution}) may also be  made by inserting  the tortoise coordinate    
 (\ref{modtortoise1})    into (\ref{modtort4}) and (\ref{modtort6}). As the limiting patterns of the solution (\ref{generalpowersolution}), obtained
in two different ways described above, must obviously match each other,
bringing them together directly fixes  the parameters $\epsilon$ and $\delta$:
\begin{equation}  \label{epsilondelta}
\delta = -i \frac{r_+^2}{r_+ - r_-} \Big( \omega - \frac{qQ}{r_+} \Big) - s, \qquad
\epsilon =   i\omega (r_+ + r_- ) -1 -2s  - i qQ .
\end{equation}
It is worthy to note  that these parameters are not affected by the spacetime noncommutativity.
 Furthermore, they coincide with the corresponding parameters obtained in the reference   \cite{Chowdhury:2018pre}.   

\subsection{Recurrence relations}

After inserting the power series solution (\ref{generalpowersolution})  with exact values for the parameters (\ref{epsilondelta}) into the equation \eqref{anz11}, we obtain the recurrence relations for the coefficients $a_n$. While in commutative case we get 3-term recurrence relations, in the noncommutative case we get the 6-term recurrence relation
\begin{eqnarray}  \label{6contfr}
A_n a_{n+1} + B_n a_n +C_n a_{n-1} + D_n a_{n-2} + E_n a_{n-3} + F_n a_{n-4 }  &=& 0, \quad  n\geqslant 4\nonumber \\
A_3 a_{4} + B_3 a_3 +C_3 a_{2} + D_3 a_{1} + E_3 a_{0}   &=& 0, \quad n=3\nonumber \\
A_2 a_{3} + B_2 a_2 +C_2 a_{1} + D_2 a_{0}   &=& 0, \quad n =2 \\
A_1 a_{2} + B_1 a_1 +C_1 a_{0}    &=& 0, \quad n=1\nonumber \\
A_0 a_{1} + B_0 a_0   &=& 0, \quad n=0.  \nonumber
\end{eqnarray}

The coefficients $A_n, B_n, C_n, D_n, E_n $ and $ F_n$ are given by

\begin{align}   
  A_n   &=   r_+^3 \alpha_{n},  \nonumber \\
&\nonumber\\
  B_n  &=  r_+^3 \beta_n - 3 r_+^2 r_- \alpha_{n-1}   \nonumber \\ 
&  \qquad -ia\nu qQr_+\Big(\frac{r_+-r_-}{2}+(n-s)(r_+-r_-)-ir_+(\omega r_+-qQ)+(r_+-r_-)\frac{s}{2}\Big), \nonumber 
&\nonumber\\ \nonumber\\
  C_n  &=  r_+^3 \gamma_n  + 3r_+ r_-^2 \alpha_{n-2}  -3r_+^2 r_- \beta_{n-1}\nonumber\\
 &+a\nu qQ\omega r_+(r_+-r_-)^3+ia\nu qQ(r_+-r_-)\big((r_+-r_-)^2+\frac{r_-}{2}(r_+-r_-)\big)\nonumber\\
 &-ia\nu qQ(r_+-r_-)^2(-1-2s-iqQ+i\omega (r_++r_-)) r_++ia\nu qQ(r_+-r_-)(2r_++r_-)\big((n-1-s)(r_+-r_-)\nonumber\\
 &-ir_+(\omega r_+-qQ)\big)+ia\nu qQsr_+(r_+-r_-)(r_++2r_-)-ia\nu qQs(r_+-r_-)(2r_++r_-)\frac{r_++r_-}{2},\nonumber \\
&   \label{contfr1} 
\end{align}

\begin{align}
  D_n  &= - r_-^3 \alpha_{n-3}  + 3r_+ r_-^2 \beta_{n-2} -3 r_+^2 r_- \gamma_{n-1}\nonumber\\
&+\frac{i}{2}a\nu qQr_+(r_+-r_-)^2-ia\nu qQ(r_+-r_-)(2r_++r_-)\big((n-2-s)(r_+-r_-)-ir_+(\omega r_+-qQ)\big)\nonumber\\
&+ia\nu qQ(r_+-r_-)^2(r_++r_-)\big(-1-2s-iqQ+i\omega (r_++r_-)\big)-ia\nu qQ(r_+-r_-)^3\nonumber\\
&-a\nu qQr_-\omega (r_+-r_-)^3+\frac{i}{2}a\nu qQs(r_+-r_-)(r_++2r_-)(r_++r_-)-ia\nu qQsr_-(r_+-r_-)(2r_++r_-) ,  \nonumber \\
&\nonumber\\
  E_n  &=   3r_+ r_-^2 \gamma_{n-2} - r_-^3 \beta_{n-3} \nonumber\\
&-ia\nu qQ(r_+-r_-)^2\frac{r_-}{2}-ia\nu qQ(r_+-r_-)^2\big(-1-2s-iqQ+i\omega (r_++r_-)\big)r_-\nonumber\\
&+ia\nu qQ(r_+-r_-)r_-\big((n-3-s)(r_+-r_-)-ir_+(\omega r_+-qQ)\big)\nonumber\\
&-ia\nu qQsr_-(r_+-r_-)(2r_++r_-)+\frac{i}{2}a\nu qQs(r_+-r_-)(r_++2r_-)(r_++r_-),
 \nonumber \\
&\nonumber\\
  F_n &= -r_-^3 \gamma_{n-3}.\nonumber \\ \nonumber
\end{align}

The coefficients $\alpha_n, \beta_n, \gamma_n$ are
\begin{align}
  \alpha_n  &=  -(n+1) \Big( r_-(n-s+1)+r_+(-n+s-1-2iqQ+2ir_+\omega) \Big), \label{contfrsimple} \\
    \beta_n  &=   -r_+\Big(\lambda_s+2n^2-4ir_+\omega(2n+1+3iqQ)+6inqQ+2n-4(qQ)^2+3iqQ-8r_+^2\omega^2+s+1 \Big)+\nonumber\\
  &\qquad +r_- \big(\lambda_s+2n(n+1+iqQ)+iqQ+s+1 \big)-2i(2n+1)r_+r_-\omega, \\
  \gamma_n &= -\Big(n+2i\big(qQ-\omega(r_++r_-)\big)\Big)\Big(n(r_--r_+)+ir_+(-2qQ+2r_+\omega+is)+r_-s\Big).
\end{align}
The first relation in (\ref{6contfr}), for $n\geqslant 4$, is
a general $6$-term recurrence relation. The remaining four relations are the indicial equations
relating the lowest order coefficients $a_n$ in the general expansion (\ref{generalpowersolution}).
They may be thought as boundary conditions for the  first relation in (\ref{6contfr}). In contrast to the commutative case \cite{Richartz:2014jla}, where we have the 3-term recurrence relations, in the NC case we have these 6-term recurrence relations. 
Due to differing orders of recursion, comparison with the commutative case in \cite{Richartz:2014jla} is not trivial, the main point being that the $a \to 0$ limit should  be taken not at the very end, but rather at some earlier step in the analysis. For more details we refer the reader to \cite{DimitrijevicCiric:2019hqq}. 

Having the recurrence relations that involve more than $3$ expansion coefficients $a_n$, as we do have here, we cannot straightforwardly apply the usual method for solving the recurrence relations \cite{gautschi1967computational}. Instead, we should first use the Gaussian elimination method to gradually reduce the initial recurrence relation from the $6$-term recurrence relation to a $3$-term recurrence relation. 
In our case the Gaussian elimination method  needs to be applied $3$ times in a row. The details of this calculation are presented in Appendix B of \cite{DimitrijevicCiric:2019hqq}. 
The initial recurrence relation can ultimately be reduced to a three-term recurrence relation, expressed as
\begin{align}
\tilde{\alpha}_0 a_{1} +\tilde{\beta}_0 a_{0} &= 0,\\
\tilde{\alpha}_n a_{n+1} +\tilde{\beta}_n a_{n}+ \tilde{\gamma}_n a_{n-1} &=0 \qquad \text{for} \qquad n\geq 1 .
\label{eq-threeterm}
\end{align}
Owing to the recursive nature of this procedure, a general analytic form for the coefficients $\tilde{\alpha}_n$, $\tilde{\beta}_n$, and $\tilde{\gamma}_n$ is unavailable. These coefficients are  complex and must be computed numerically.

The convergence of the series $\sum_{n}a_n$ is essential for ensuring that the ansatz \eqref{generalpowersolution} exhibits the appropriate asymptotic behavior. For convergence to hold, the coefficients $\tilde{\alpha}_n$, $\tilde{\beta}_n$, and $\tilde{\gamma}_n$ must satisfy the following continued fraction condition \cite{Leaver:1985ax}
\begin{eqnarray}
\tilde{\beta}_0 &=& \frac{\tilde{\alpha}_0 \tilde{\gamma}_{1}}{\tilde{\beta}_{1}-\frac{\tilde{\alpha}_{1}\tilde{\gamma}_{2}}{\tilde{\beta}_{2} - \cdots}} = \frac{\tilde{\alpha}_0\tilde{\gamma}_{1}}{\tilde{\beta}_{1}-}\frac{\tilde{\alpha}_{1}\tilde{\gamma}_{2}}{\tilde{\beta}_{2} -}\frac{\tilde{\alpha}_{2}\tilde{\gamma}_{3}}{\tilde{\beta}_{3} -}\cdots 
\label{eqLeaver}
\end{eqnarray}
%
The QNM frequencies are determined by finding the roots of the continued fraction \eqref{eqLeaver}. However, as the continued fraction represents an infinite series, it must be truncated to a finite number of terms to facilitate numerical evaluation. In our numerical computations, we employed approximately $N \sim 200$ terms, depending on the required precision (up to six decimal places or more). As a validation measure, we confirmed that increasing the number of terms consistently improves the accuracy of the computed QNMs.

To enhance accuracy, we apply Nollert's improvement method to estimate the contribution from the truncated tail of the series \cite{Nollert:1993zz}. The Nollert method involves recognizing that the quantity $R_N=-a_{N+1}/a_N$ satisfies the recursion relation
\begin{equation}
\label{nolrel}
R_N=\frac{\tilde{\gamma}_{N+1}}{\tilde{\beta}_{N+1} - \tilde{\alpha}_{N+1}R_{N+1}},
\end{equation}and accurately represents the remainder of the continued fraction \eqref{eqLeaver} truncated at order $N$. This remainder $R_N$ is subsequently expanded in terms of inverse powers of $N^{1/2}$ as follows
\begin{equation}
\label{nolexpan}
R_N= \sum_{k=0}^{\infty} C_k N^{-k/2}.
\end{equation}
Substituting this series expansion into equation \eqref{nolrel} allows one to determine the coefficients $C_k$. Since analytic expressions for $\tilde{\alpha}_{N+1}$, $\tilde{\beta}_{N+1}$, and $\tilde{\gamma}_{N+1}$ are not feasible (they are evaluated numerically using the Gaussian elimination procedure), we rely on the commutative expressions for $C_k$ provided in~\cite{Richartz:2014jla} \footnote{The Nollert method enhances the convergence of the continued fraction equation. Since the commutative tail approximation suffices for our analysis, we adopt this approach in our study.}.

Using described  methods and corresponding inputs, we employ a numerical root-finding algorithm to calculate the noncommutative QNM frequencies, whose results are presented in the subsequent section.



 \section{Quasinormal-mode spectrum} \label{results}

\begin{figure}
\centering
	\subfigure[]{\includegraphics[width=0.48\textwidth]{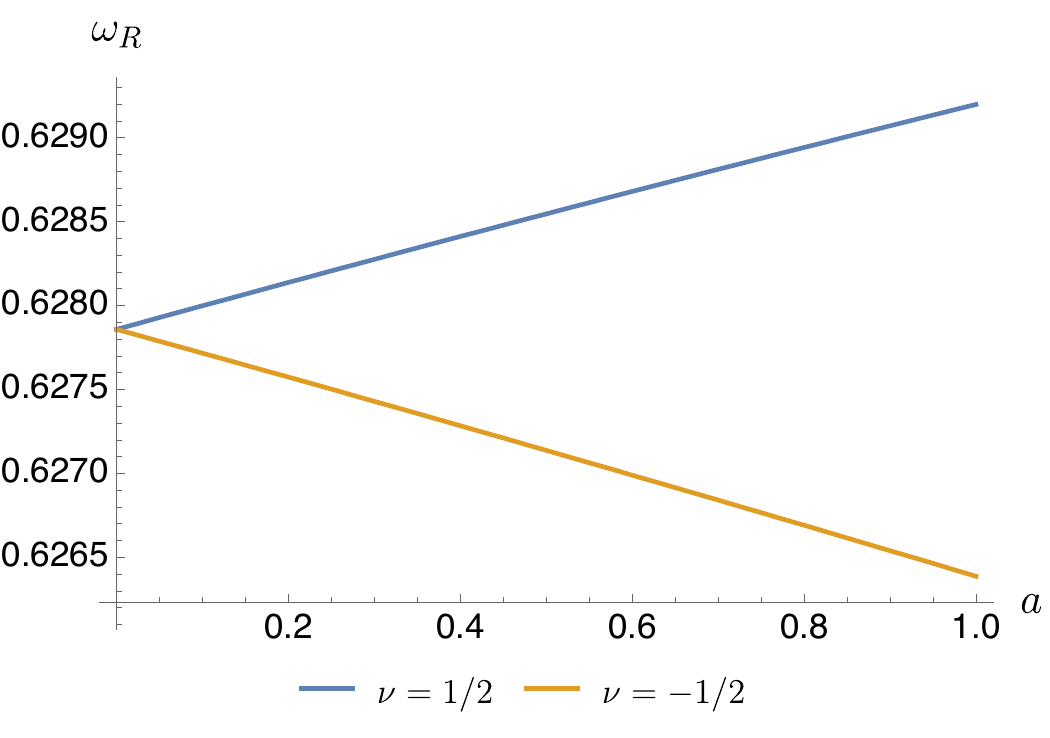}\label{fig1a}} 
\quad
	\subfigure[]{\includegraphics[width=0.48\textwidth]{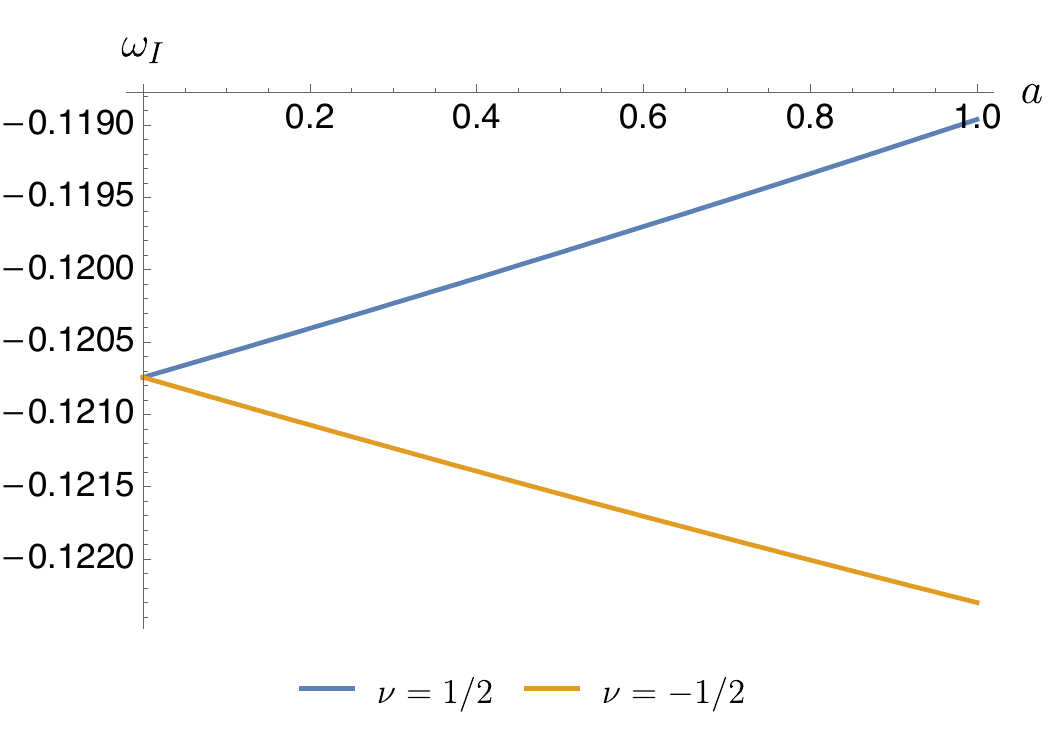}\label{fig1b}} 
    \subfigure[]{\includegraphics[width=0.48\textwidth]{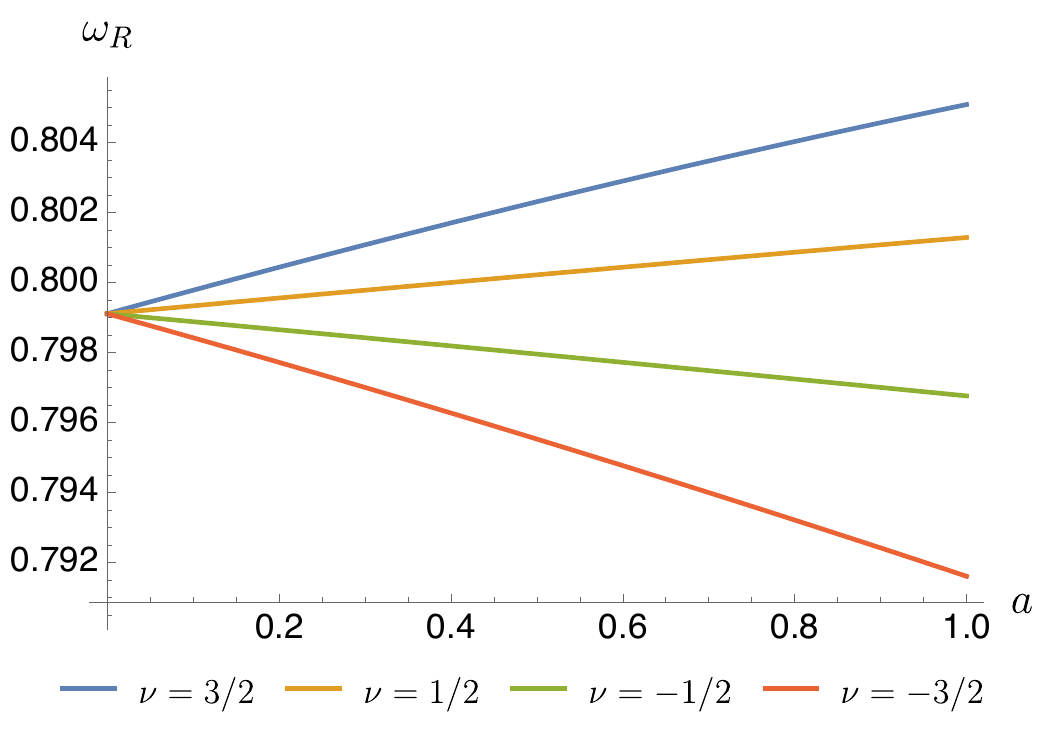}\label{fig2a}} 
\quad
	\subfigure[]{\includegraphics[width=0.48\textwidth]{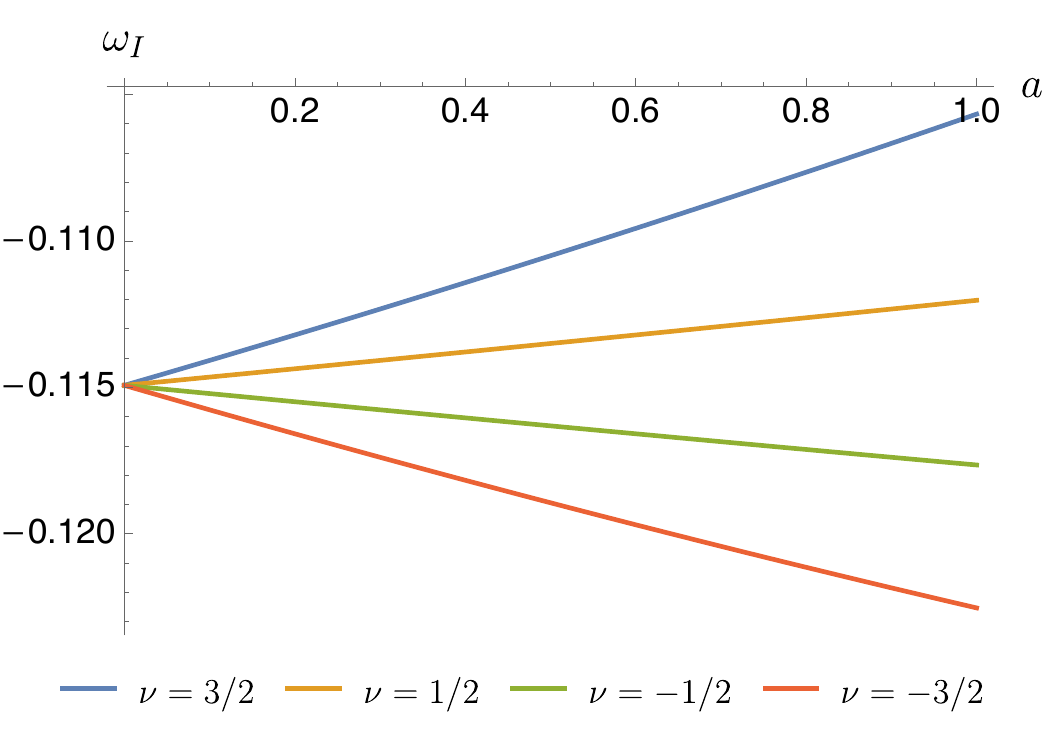}\label{fig2b}} 
	\caption{Dependence of the fermionic QNM frequencies on the NC parameter $a$ for $j=1/2$ and $s=1/2$. Subfigure (a) displays the splitting in the real part of QNM frequencies, while subfigure (b) illustrates the corresponding splitting in the imaginary part for various values of the magnetic quantum number $\nu$. Subfigures (c) and (d) display the spectrum for $j = 3/2$ and $s = 1/2$. The parameters are set to $Q = 0.5$, $qQ = 1$, and $M = 1$.} 
\label{fig1}
\end{figure}

 The resulting spectrum displays interesting deviations from the commutative one. Figure \ref{fig1} displays the dependence of the fundamental mode QNM frequency on the NC parameter $a$. Linear deviation from the commutative frequency supports the validity of continuous fraction method for obtaining the spectrum in this perturbative approach. This characteristic Zeeman-like splitting is similar to the scalar field case considered in \cite{DimitrijevicCiric:2019hqq} and spin-2 field analysed in \cite{Herceg:2023pmc,Herceg:2023zlk,Herceg:2024vwc}\footnote{There it should be compared to $r - \varphi$ noncommutativity as the authors do not consider the $t - \varphi$ noncommutativity studied here.}. It should be noted that the expected value of $a$ is at the order of Planck Length, which is much lower than 1 in units where the black hole mass is equal to 1. Therefore the range of values on the horizontal axis of Figure \ref{fig1} is selected for illustration purpose.

\begin{figure}
\centering
	\subfigure[]{\includegraphics[width=0.48\textwidth]{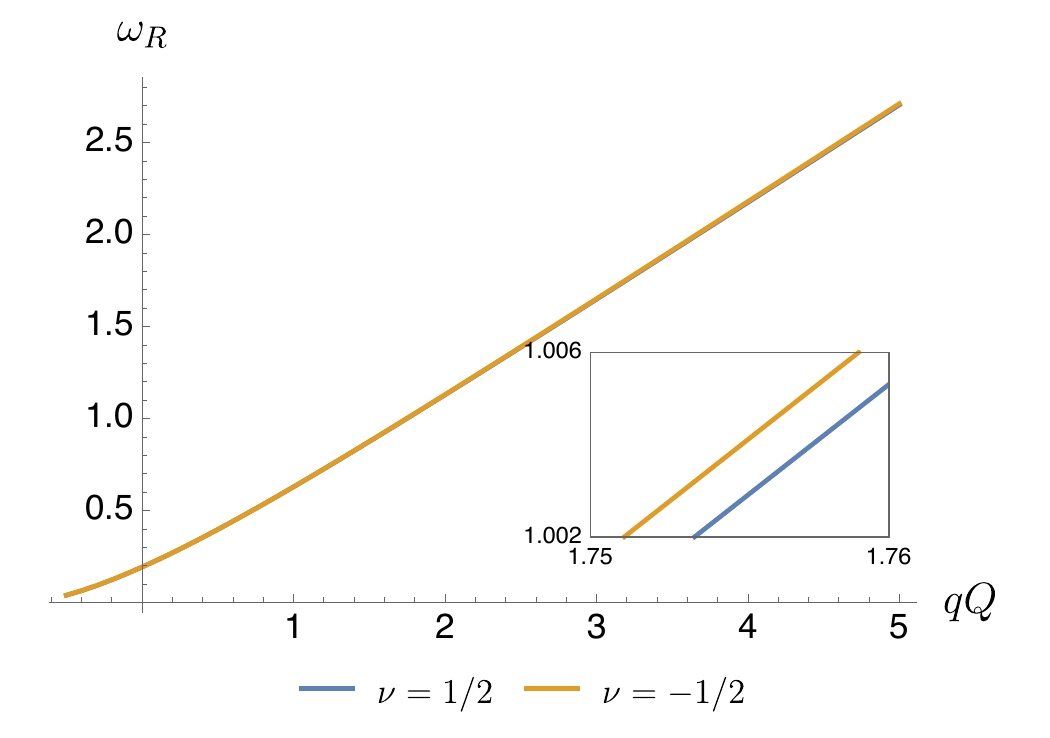}\label{fig4a}} 
\quad
	\subfigure[]{\includegraphics[width=0.48\textwidth]{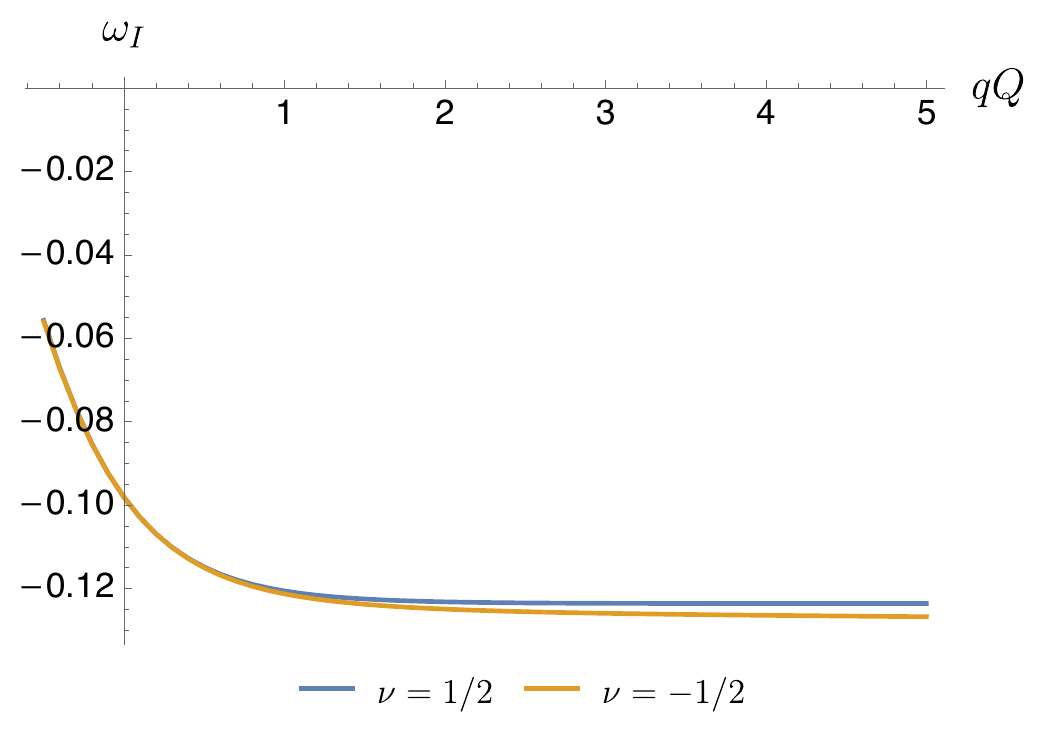}\label{fig4b}} 
    \subfigure[]{\includegraphics[width=0.48\textwidth]{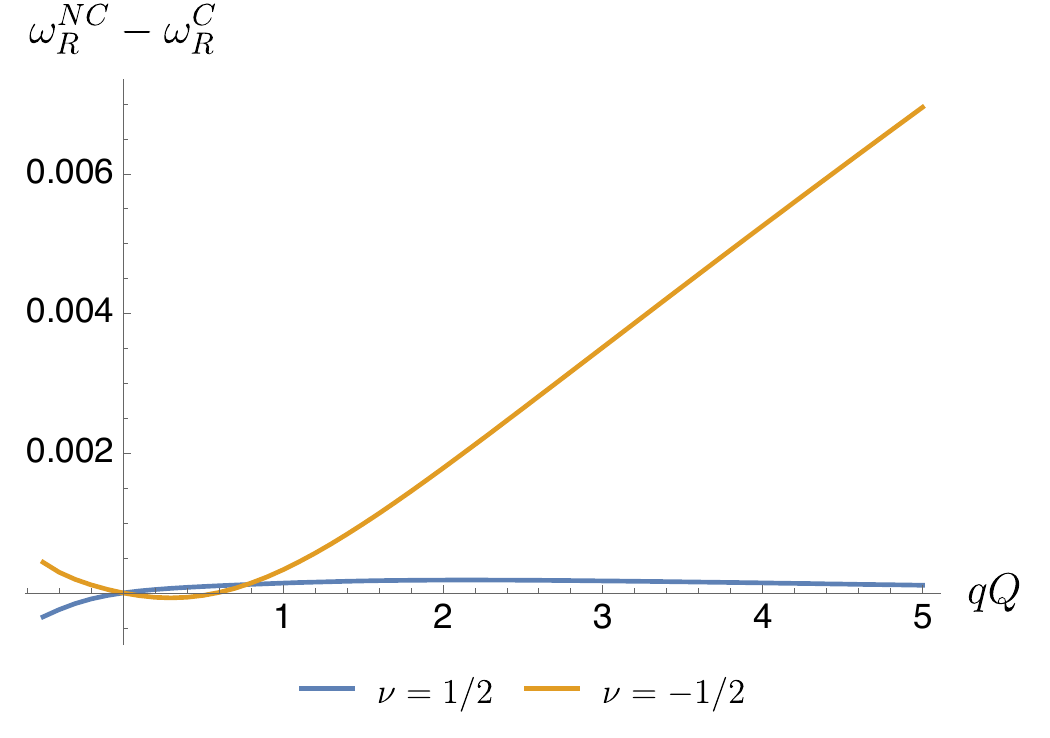}\label{fig4c}} 
\quad
	\subfigure[]{\includegraphics[width=0.48\textwidth]{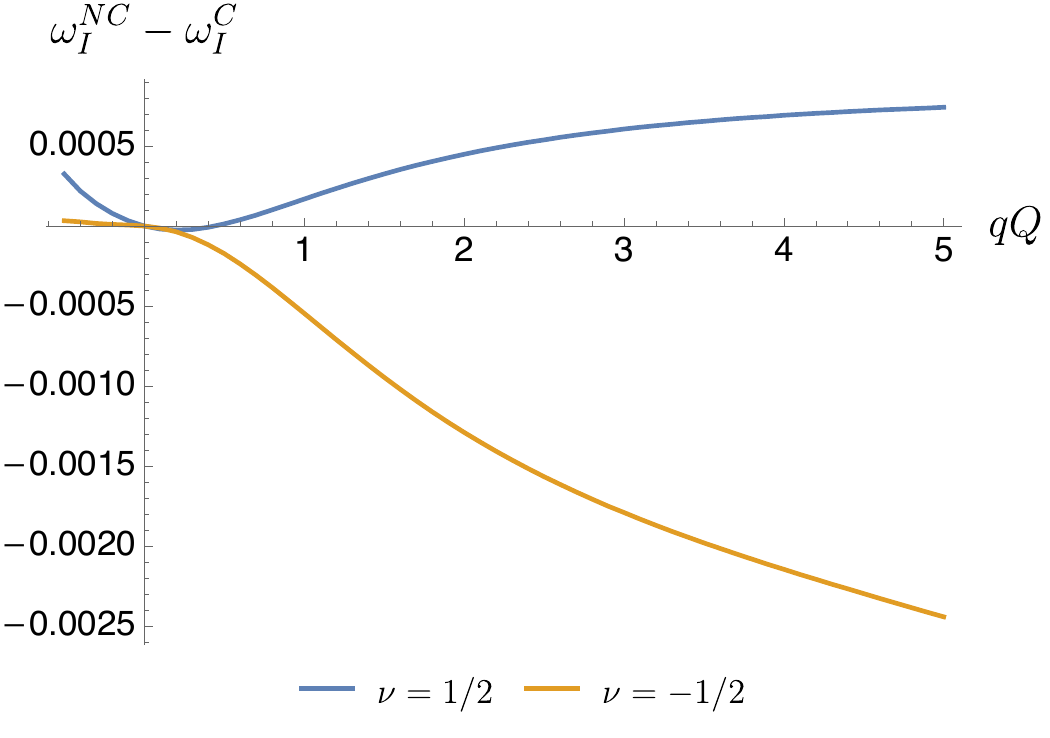}\label{fig4d}} 
	\caption{Dependence of fermionic QNMs ($j=1/2$, $s=1/2$) on the parameter $qQ$. Subfigures (a) and (b) respectively illustrate the splitting in the real (highlighted in the inset) and imaginary parts of the QNM frequencies for different magnetic quantum numbers $\nu$. Subfigures (c) and (d) display the difference between NC and commutative QNM frequencies, $\omega^{NC}-\omega^{C}$, as functions of $qQ$ for the real and imaginary parts, respectively. Parameters used are $Q = 0.5$, $a = 0.1$, and $M = 1$.} 
\label{fig4}
\end{figure}

In Figure \ref{fig4} the QMM spectrum is plotted as a function of $q Q$ , while keeping $Q$ fixed. The impact of noncommutativity on the imaginary part of the frequency (damping part) appears to be greater than the impact on the real part, especially for higher values of $q Q$. However, when one takes into account the scales on the y-axis, the differences are comparable in absolute value.
A similar graph for the scalar field can be found in \cite{DimitrijevicCiric:2019hqq}. The trend is very similar, although the spectrum of the scalar field seems to be affected more by spacetime noncommutativity.

In (c) and (d) subfigures of Figure \ref{fig4}, it is visible that the deviation of $\nu = -1/2$ mode from the commutative value is much larger than that of $\nu = 1/2$ mode -- especially in the real part.
While this behaviour looks interesting, the qualitative difference between the QNM frequencies for $\nu = \pm1/2$ (which does not exist in commutative case \cite{Richartz:2014jla}) could be explained by preferred direction of $\varphi$ in the twist \eqref{AngTwist0Phi}. Another possible source of asymmetry  could be that the two components of the chirality display different behaviour and our analysis of the spectrum was concerned only with the $s=+1/2$ component of the spinor.

\begin{figure}
\centering
	\subfigure[]{\includegraphics[width=0.48\textwidth]{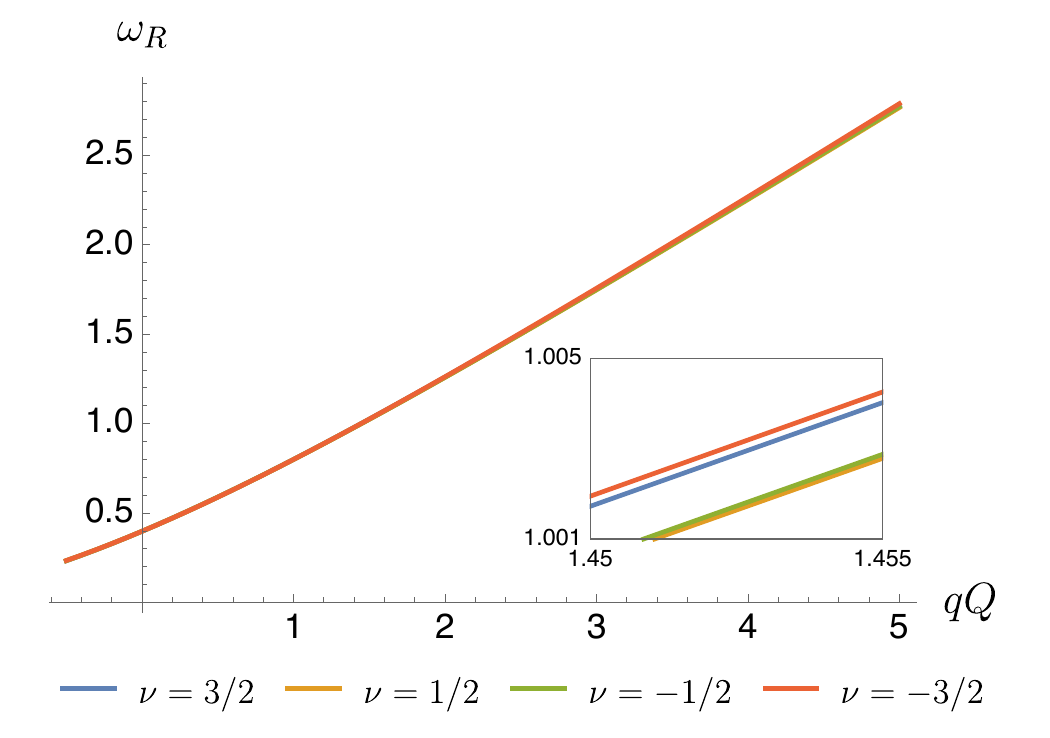}\label{fig5a}} 
\quad
	\subfigure[]{\includegraphics[width=0.48\textwidth]{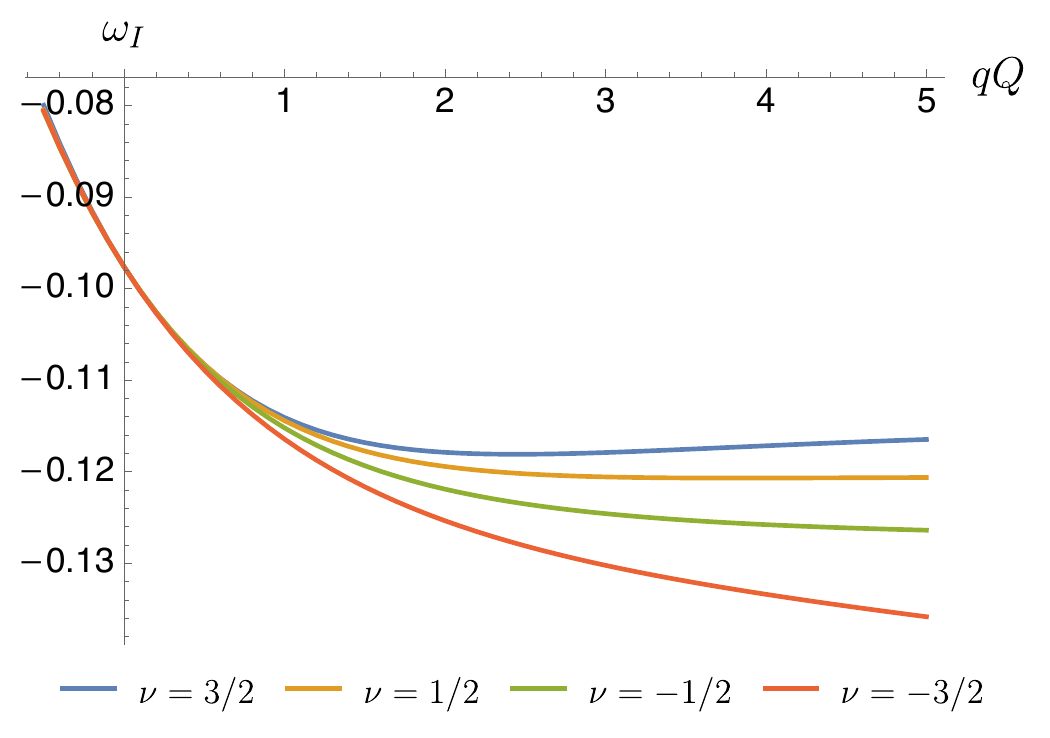}\label{fig5b}} 
    \subfigure[]{\includegraphics[width=0.48\textwidth]{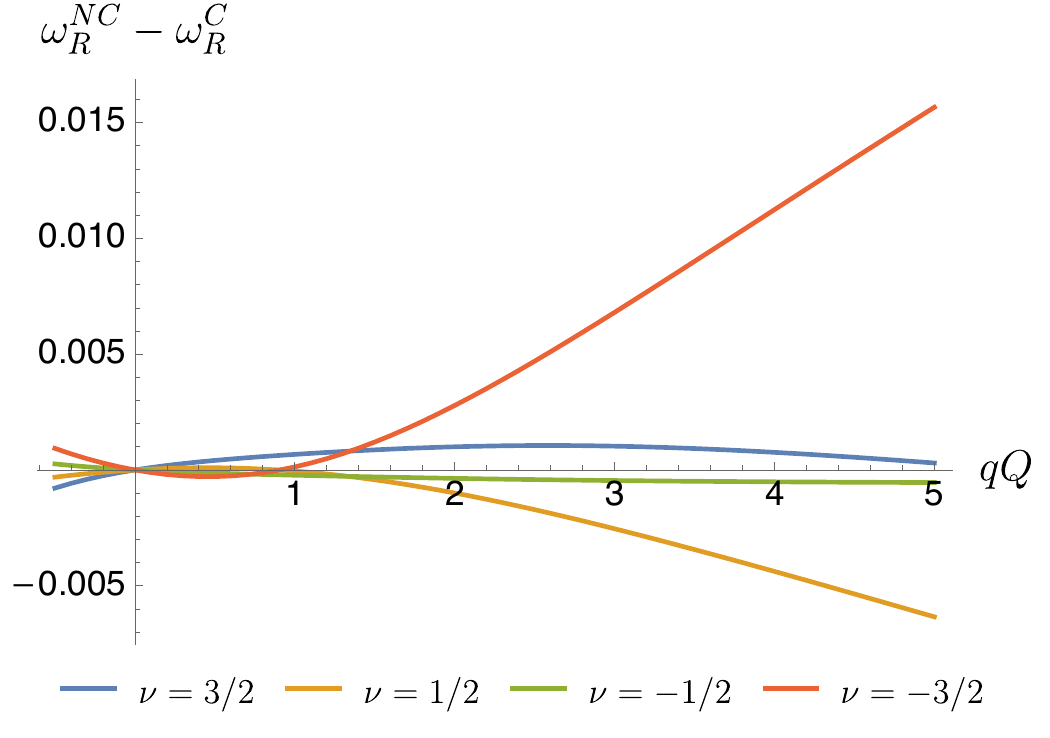}\label{fig5c}} 
\quad
	\subfigure[]{\includegraphics[width=0.48\textwidth]{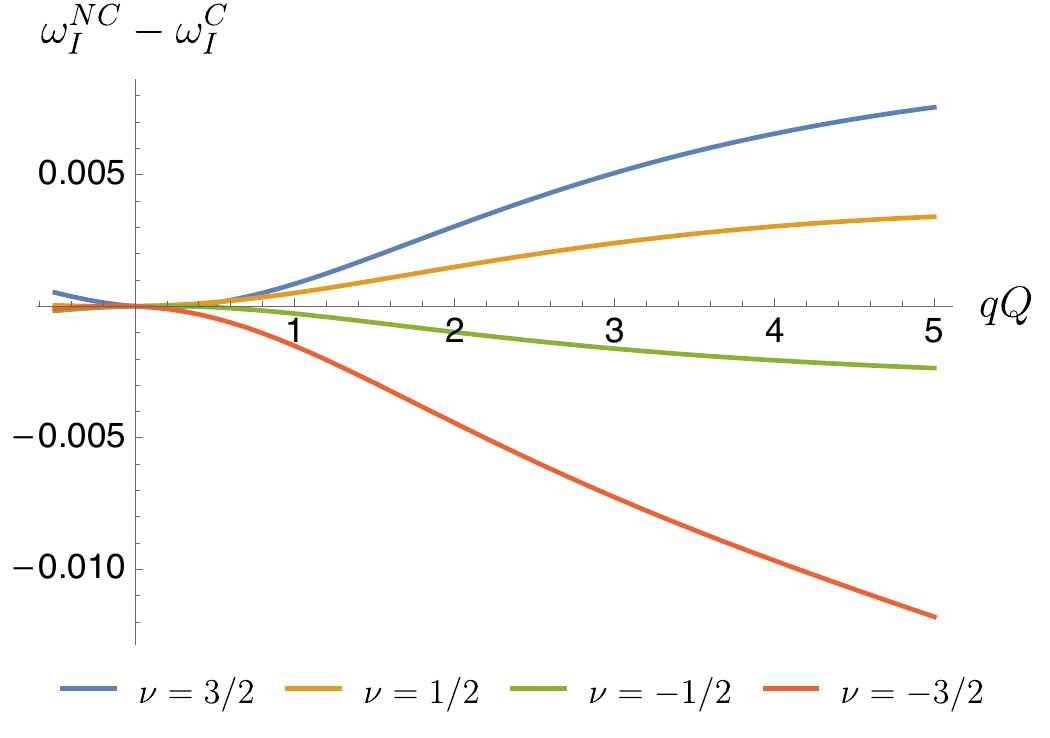}\label{fig5d}} 
	\caption{Dependence of fermionic QNMs ($j=3/2$, $s=1/2$) on $qQ$. Subfigures (a) and (b) respectively illustrate the splitting in the real (highlighted in the inset) and imaginary parts of QNM frequencies for various magnetic quantum numbers $\nu$. Subfigures (c) and (d) depict the differences between NC and commutative QNM frequencies, $(\omega^{NC}-\omega^{C})$, as functions of $qQ$ for the real and imaginary parts, respectively. Parameters chosen are $Q=0.5$, $a=0.1$, and $M=1$.} 
\label{fig5}
\end{figure}

In Figure \ref{fig5} the fermionic QNM spectrum of the fundamental mode is plotted for $j = 3/2$ and accordingly four values of $\nu$. In the subfigure \ref{fig5b} one can observe an almost uniform splitting of imaginary parts of the frequencies. 
Even though the additional component in the modified Reissner-Nordstr\"om metric \eqref{NCdsRN} cannot be related to rotation, this kind of splitting of imaginary part of frequencies temptingly
points to such interpretation. Indeed, as imaginary parts represent damping, we may infer from the figures that the waves with negative value of $\nu$ are damped more strongly than the ones with positive $\nu$.
This is reminiscent of a way how frame dragging advances the waves orbiting in the same direction as the black hole rotates and damps the ones rotating in the opposite direction \cite{Detweiler:1980gk, Leaver:1985ax}. The subfigures \ref{fig5c} and \ref{fig5d} show the noncommutative deviations in the real and imaginary parts of the frequencies, respectively.

\begin{figure}
\centering
	\subfigure[]{\includegraphics[width=0.45\textwidth]{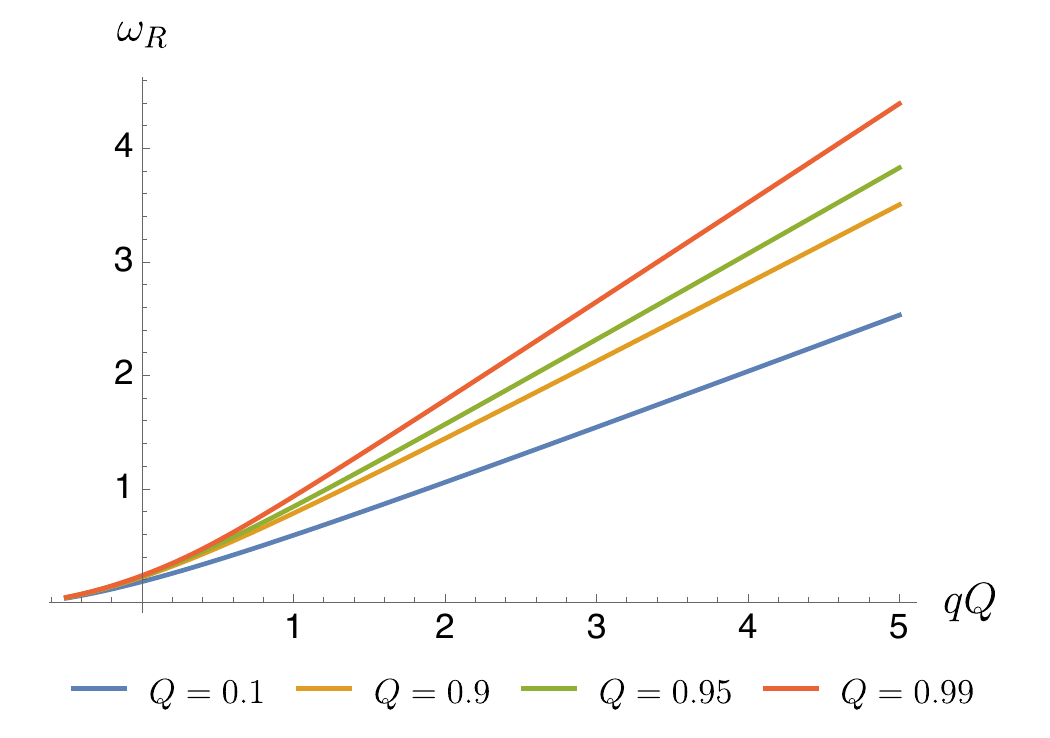}\label{fig6a}} 
\quad
	\subfigure[]{\includegraphics[width=0.45\textwidth]{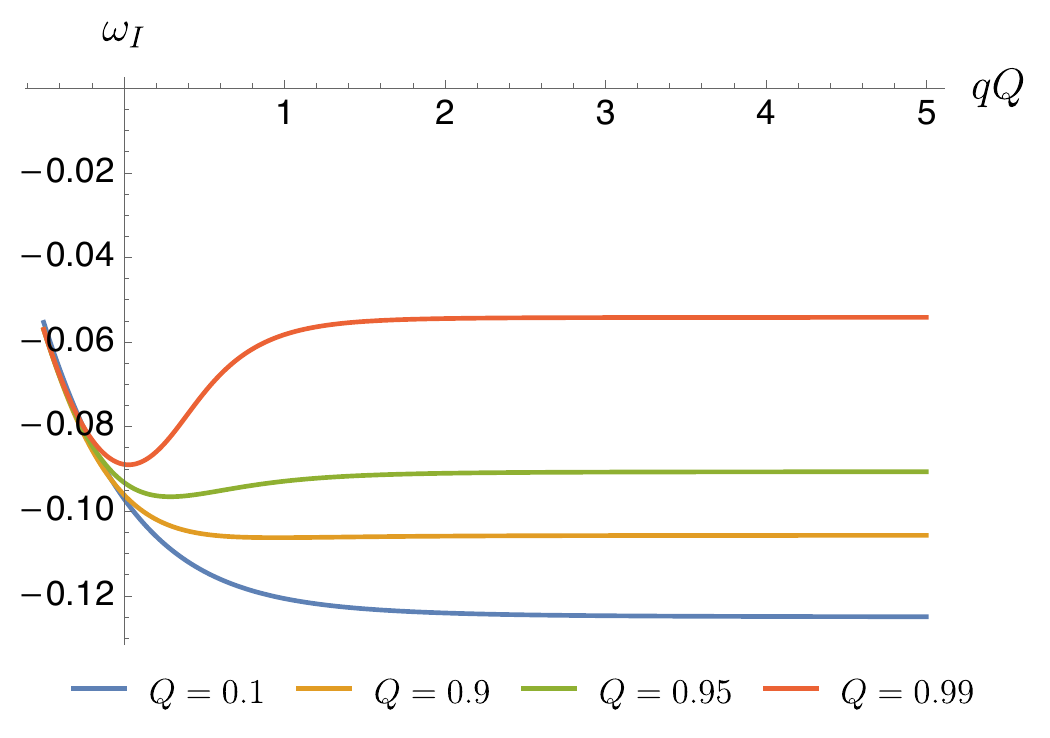}\label{fig6b}} 
    \subfigure[]{\includegraphics[width=0.45\textwidth]{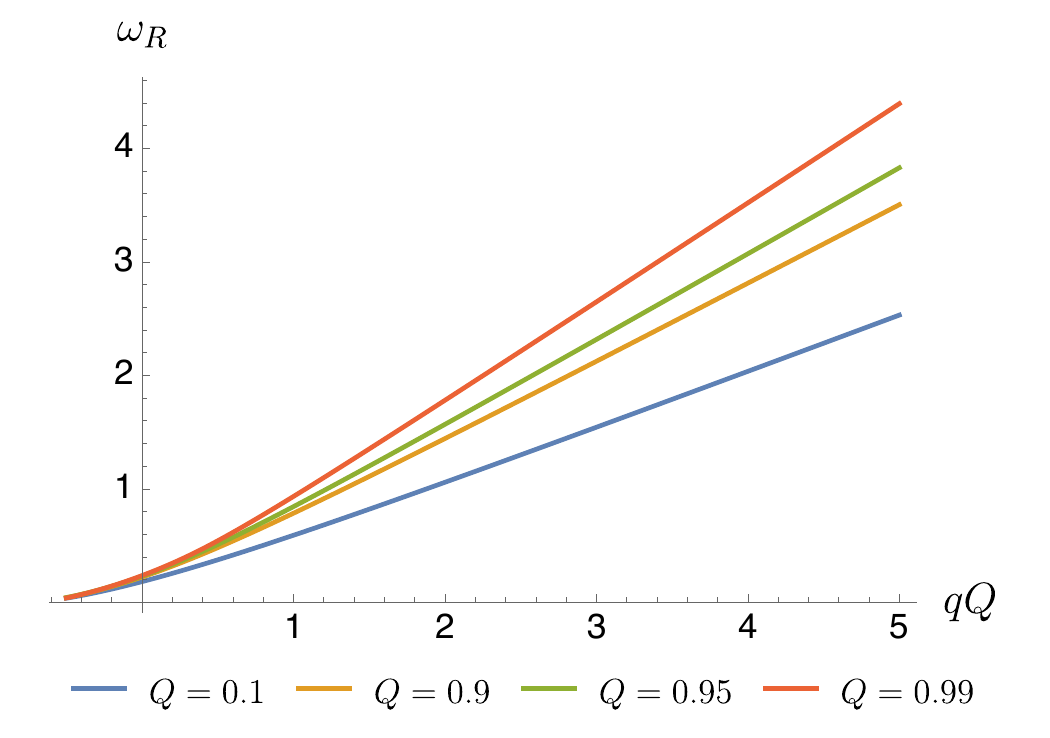}\label{fig6c}} 
\quad
	\subfigure[]{\includegraphics[width=0.45\textwidth]{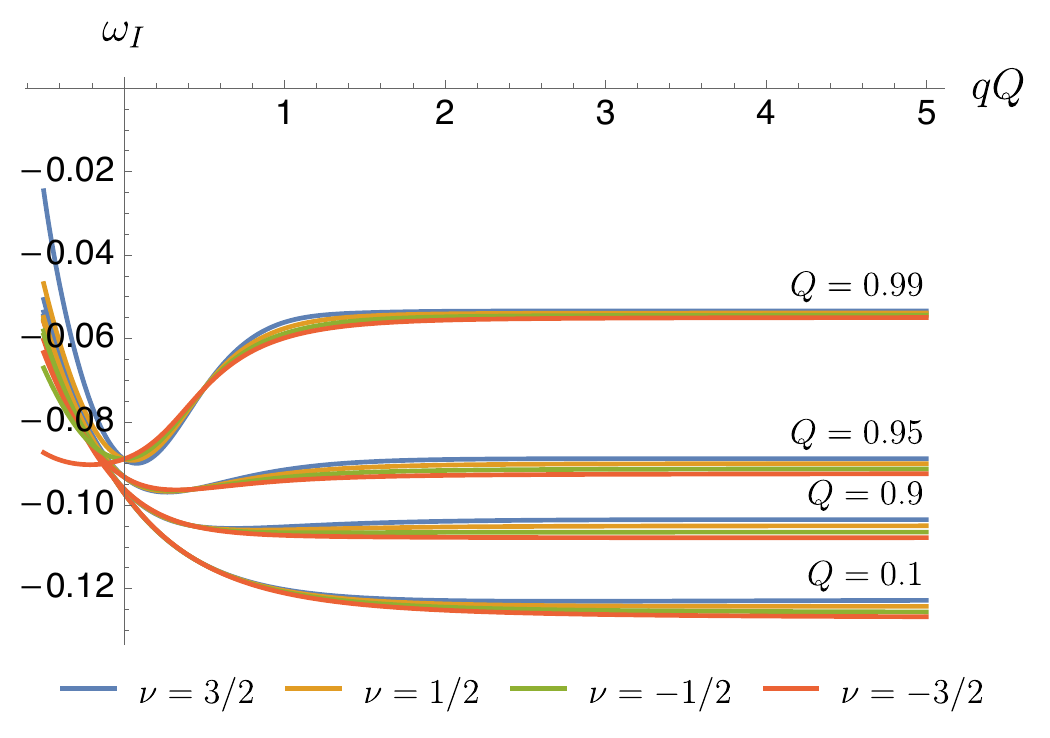}\label{fig6d}} 
	\caption{Variation of fermionic QNMs ($j=3/2$, $s=1/2$) against the parameter $qQ$ for different values of $Q$. Subfigures (a) and (b) respectively illustrate the real and imaginary parts of the QNM frequencies in the commutative case. Subfigures (c) and (d) represent the NC QNM frequencies (with $a=0.1$) plotted for $j = 3/2, s = 1/2$ and $\nu \in \{-3/2, -1/2, 1/2, 3/2\}$. The splitting in $\nu$ is not visible in the (c) subfigure, but is clearly present in (d). For both commutative and NC cases, we set $M=1$.}
\label{fig6}
\end{figure}

Figure \ref{fig6} similarly shows the $qQ$ dependence of the QNM frequencies for several values of $Q$. The most interesting subfigure is \ref{fig6d}, where one can see how much noncommutativity affects the spectrum depending on the value of $qQ$ for fixed $Q$-values. As mentioned previously, the commutative spectrum does not depend on $\nu$, so plotting for $\nu$ from $-3/2$ to $3/2$ provides an insight into the magnitude of the NC correction to the spectrum.

From subfigure \ref{fig6b} one may observe a phase transition-like appearance in the behaviour of the $ \omega_I$ plot occurring at around $Q/M \sim 0.9$. This threshold is not significantly affected by noncommutativity \cite{Richartz:2014jla}. For the scalar NC case, the phase transition occurs at $Q/M \sim 0.7$ \cite{DimitrijevicCiric:2019hqq}.

\begin{figure}
\centering
	\subfigure[]{\includegraphics[width=0.45\textwidth]{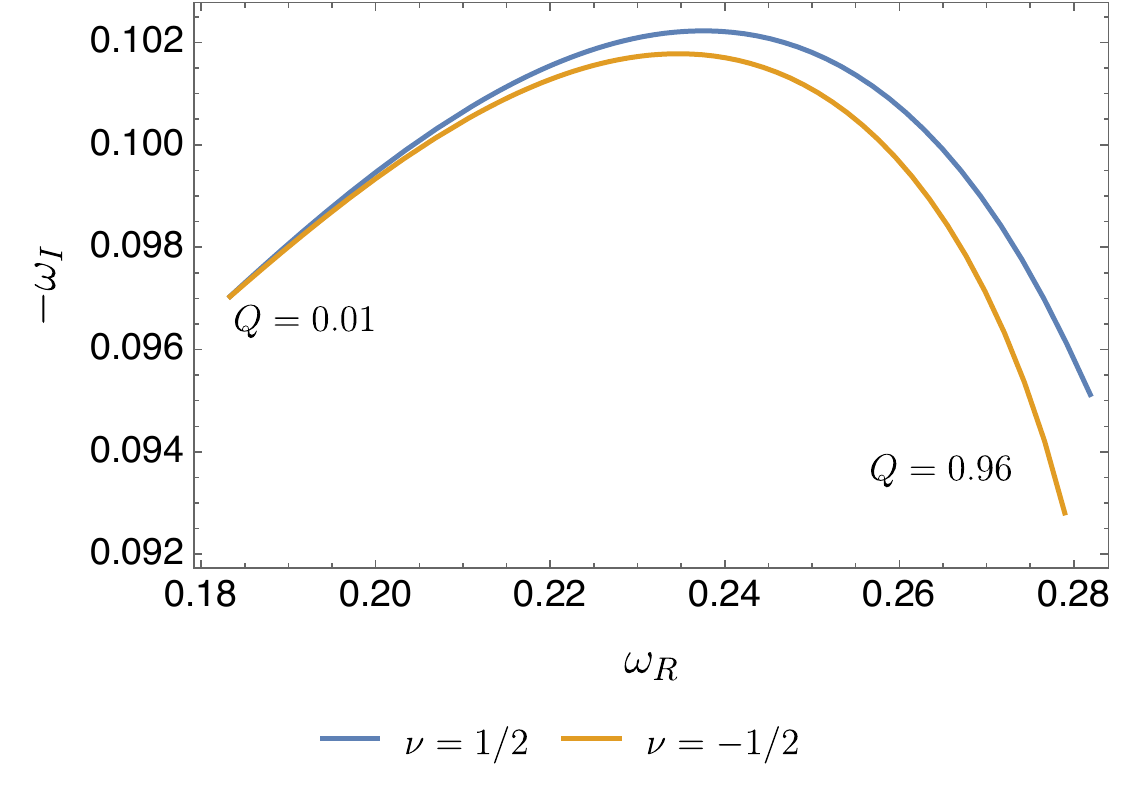}\label{fig7a}} 
\quad
	\subfigure[]{\includegraphics[width=0.45\textwidth]{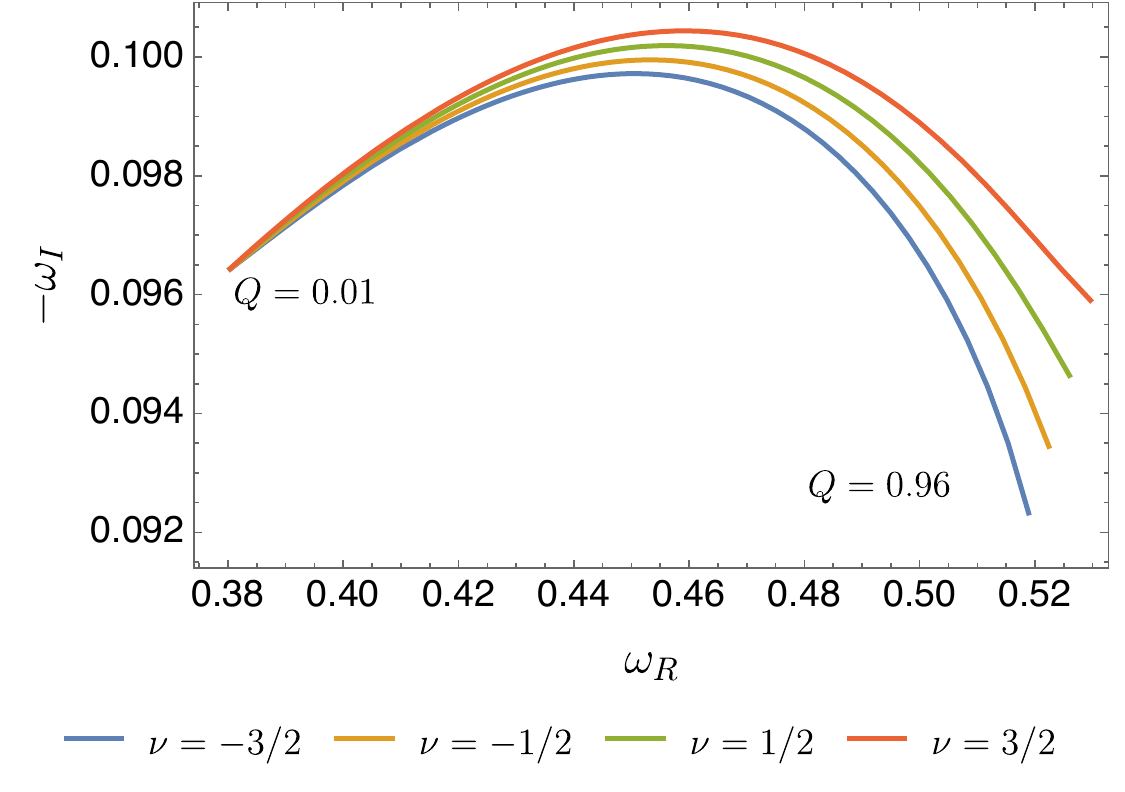}\label{fig7b}} 
	\caption{Real and imaginary parts of fermionic QNMs for different magnetic quantum numbers $\nu$ with $Q$ varying from $0.01$ to $0.96$. Subfigure (a) corresponds to the case $(j=1/2, s=1/2)$, and subfigure (b) corresponds to $(j=3/2, s=1/2)$. For both cases, we set parameters as $q=0.1$, $a=1$, and $M=1$. }
\label{fig7}
\end{figure}

Figure \ref{fig7} shows the profile of the fundamental QNM in the complex plane, where the frequency plot is parametrized by $Q/M$. 
The impact of noncommutativity grows with increasing $Q/M$. The characteristic shape of the graph is not significantly affected by noncommutativity. 
Note that at values of $Q/M$ at around $0.96$ the frequencies  start to diverge more rapidly.

Our analysis breaks down at higher values of the black hole charge \( Q \). The exact threshold of \( Q \) where this occurs depends additionally on parameters such as the field charge \( q \) and the noncommutative parameter \( a \). Nevertheless, for very small values of \( a \), our method remains robust for arbitrary choices of other parameters, except in the near-extremal limit \( Q/M \to 1 \). Since the magnitude of \( a \) is expected to be on the order of the Planck length, the regime of validity for our analysis essentially mirrors that of the commutative case. As in the commutative scenario, the extremal limit necessitates a separate, dedicated treatment.


\section{Final remarks} \label{conclusion}

By the progress in experimental astrophysics and discovery of gravitational
waves, black hole QNMs became one of the most important  tools for testing  predictions of general relativity  in hitherto unexamined   conditions,  as well as for  gaining insights into the characteristic parameters of massive compact objects  \cite{Berti:2009kk}. The latter may be mainly achieved through the study of quasinormal mode spectra of black holes and other exotic compact objects.

Studying the black hole perturbations  by different types of fields gives rise to the details in the spectrum that  are able to uncover  features like   instabilities, bound states,
zero modes or characteristic resonances  for which perturbations of these  field  configurations tend to be radiated   away and decay. The spectrum may also tell us a lot about the properties  of the spacetime itself. In particular,
if the structure of the spacetime is not in line with the usual concept of smooth continuum, then presumably this should be also reflected in the spectrum.

  In order to emulate  conditions like this, in this paper we have used a specific model in which a massless charged Dirac field  probes a modified  Reissner–Nordstr\"om geometry. This physical model  was previously shown to be equivalent to the noncommutative gauge field theory where the NC Dirac field probes classical Reissner–Nordstr\"om  background \cite{DimitrijevicCiric:2022ohs}.

Features like these show that the total impact of  spacetime deformation may, at the level of effective description, be squeezed  into a modification of  geometry, which manifests in the appearance of an additional nonvanishing component in the original metric. This results with modified geometry of  Reissner–Nordstr\"om   type with the emergence of an additional  $r - \varphi$ component in the metric. We then took  advantage of the stated correspondence in order to analyse the modifications in the fermion perturbation spectrum that appear due to spacetime deformation.

In brief, the present work discusses the perturbations of  massless charged Dirac particles and the fermion QNMs generated by these perturbations in the vicinity of a quantum-deformed Reissner–Nordstr\"om black hole. In order to accomplish this task, we have adopted the continued fraction method extended to include the Gaussian elimination procedure. The characteristic pattern of the fundamental mode has been analysed in dependence of different black hole parameters, such as charge, mass and the NC deformation parameter. Moreover, 
changes in the spectrum were analysed for different extremalities, i.e. for different  mass over charge ratios and for different charges of the probing fermion field. 

The feature that is the most striking one and that is at the same time the only genuinely intrinsic  to noncommutative nature of space is the Zeeman-like splitting in the spectrum. The latter occurs with varying a projection of the total angular momentum. Another noteworthy feature is the dependence of the imaginary (damping) part of the frequency on $\nu$. Even though the new nonzero component in the modified RN metric cannot be directly linked to a rotation, the pattern of how the imaginary part of the frequency depends on $\nu$ gives a spectral analogue of the rotation-induced dynamics, thus providing some arguments in support of the rotational correspondence.

The study carried out in this work leaves ample room for several additional directions of research. For instance, the question of how spacetime deformation influences the QNM spectrum for massive fermion perturbations remains unanswered.  In general, it is known that the perturbations of the 
massive Dirac field give rise to the real part of the QNM frequency that increases with the mass. Likewise, the imaginary part  which corresponds to damping   decreases, implying that the QNM frequencies due to perturbations of black hole background with more massive Dirac fields
are  more likely to be detected  \cite{Chakrabarti:2008xz}. This is due to  the fields with higher
masses decaying more  slowly. How the deformed structure of spacetime would influence these certainties still remains to be seen.
Finally, it is noteworthy that Dirac fields  have been extensively studied in a  number of black hole models and gravitational theories and in a  variety of  physical processes that occur near the black hole horizon, such as the scattering and absorption processes for Dirac particles \cite{Gaina:1983ar}, the spectral power emission of Dirac fermions \cite{Page:1977um,Gaina:1985xg}, including the study of superradiance \cite{Lee:1977gk} and gray body factors \cite{Aharony:1999ti, Kanti:2014dxa, Oh:2008tc, Sakalli:2022xrb}.

A clear picture about the  influence of  quantum structure of spacetime on majority of these features and processes is still lacking as the current investigation has  only just began to scratch the surface of a   fundamental mechanisms that lie behind them.
Some of these issues, as seen from the perspective of quantum gravity, particularly massive Dirac perturbations and gray body factors corresponding to fermion particles, will be addressed in the upcoming work. \\


\noindent{\bf Acknowledgment}\\
  This  research was supported by the Croatian Science
Foundation Project No. IP-2020-02-9614 {\it{Search for Quantum spacetime in Black Hole QNM spectrum and Gamma Ray Bursts. The work of N.K.  is supported by Project 451-03-136/2025-03/200162 of the Serbian Ministry of Science, Technological Development and Innovation. }}

\bibliography{BibTex}

\end{document}